\documentclass{article}
\usepackage{fortschritte}
\usepackage{epsf}
\usepackage{amssymb,amsmath}
\usepackage{slashed}
\usepackage[nosort]{cite}

\def\be{\begin{equation}}                     %
\def\ee{\end{equation}}                       %
\def\bea{\begin{eqnarray}}                     
\def\eea{\end{eqnarray}}                       
\def\Tr{\mathop{\mathrm{Tr}}}
\def\nn{\nonumber}

\newcommand\eqn[1]{(\ref{#1})}

\newcommand{\ft}[2]{{\textstyle\frac{#1}{#2}}}
\newcommand{\cO}{{\cal O}}

\newcommand{\mb}{\mathversion{bold}}
\newcommand{\mn}{\mathversion{normal}}

\newcommand\gym{g_{\rm{YM}}}
\newcommand\vev[1]{\langle\, #1  \,\rangle}
\newcommand\MMvev[1]{\langle\, #1  \,\rangle_{\rm M\, M}}

\newcommand\entspricht{\mathrel{\hat{=}}} 
\newcommand{\Z}{\mathbb{Z}}                                
\newcommand{\R}{\mathbb{R}}                               
\newcommand{\unit}{\mathbf{1}}  
\newcommand{\textindex}[1]{{\scriptscriptstyle\mathrm{#1}}}

\def\moth{\mathsurround=0pt}
\newdimen\zo \zo=0pt

\def\tick{\leaders\hrule height 0.5ex depth 0pt \hskip 0.5pt}
\def\upboxfill{$\moth \setbox\zo\hbox{\tick}%
  \hskip 2pt\hbox to 0pt{$\tick$\hss}\hrulefill \hbox to 6pt{$\tick$\hss}$}
\def\underbox#1{\offinterlineskip{\mathord{\mathop{\vtop{\moth\ialign{##\crcr
      $\hfil\displaystyle{#1}\hfil$\crcr\noalign{}
      {\upboxfill}\crcr\noalign{}}}}\limits}}}
\def\dtick{\leaders\hrule height .34pt depth .5ex \hskip 0.5pt}
\def\downboxfill{$\moth \setbox\zo\hbox{\dtick}%
  \hskip 2pt\hbox to 0pt{$\dtick$\hss}\hrulefill \hbox to 6pt{$\dtick$\hss}$}

\def\undersym#1{\underbox{{}#1}}

\begin {document}                 
\rightline{AEI-2003-061}
\rightline{hep-th/0307101}
\def\titleline{
Lectures on the \\[0.5cm]Plane--Wave String/Gauge Theory Duality}
\def\email_speaker{{\tt 
plefka@aei.mpg.de
}}
\def\authors{
Jan Christoph Plefka\footnote{Lectures given at the RTN Winter
School on Strings, Supergravity and Gauge Theory, Turin 7-11 January 2003.}}
\def\addresses{
Max-Planck Institute for Gravitational Physics\\
(Albert-Einstein-Institute)\\
Am M\"uhlenberg 1, D-14476 Potsdam, Germany\\[0.3cm]
{\tt plefka@aei.mpg.de}
}
\def\abstracttext{
These lectures give an introduction to the novel duality
relating type IIB string theory in a 
maximally supersymmetric plane-wave background to ${\cal N}=4$, $d=4$, 
$U(N)$ Super Yang-Mills theory in a particular large $N$ and large R-charge 
limit due to Berenstein, Maldacena and Nastase.
In the first part of these lectures
the duality is derived from the AdS/CFT correspondence
by taking a Penrose limit of the $AdS_5\times S^5$ geometry and
studying the corresponding double-scaling limit on the gauge theory 
side. The resulting free plane-wave superstring is then
quantized in light-cone gauge.  On the
gauge theory side of the correspondence
the composite Super Yang-Mills operators 
dual to string excitations are identified, and it is shown how
the string spectrum can be
mapped to the planar scaling dimensions of these
operators.
In the second part of these lectures
we study the correspondence at the interacting respectively non-planar
level.
On the gauge theory side
it is demonstrated that the large $N$ large R-charge limit in 
question preserves contributions from Feynman graphs of all genera
through the emergence of a new genus counting parameter --
in agreement with the string genus expansion for non-zero $g_s$.
Effective quantum mechanical tools to compute higher genus contributions
to the scaling dimensions of composite operators are developed and
explicitly applied in a genus one computation. 
We then turn to the interacting string theory
side and give an elementary introduction into light-cone
superstring field theory in a plane-wave background and point out
how the genus one prediction from  gauge theory can be
reproduced. Finally, we summarize the 
present status of the 
plane-wave string/gauge theory duality.
}

\large
\makefront

\newpage


\section{Introduction}

Ever since the work of 't Hooft in 1974  \cite{thooft1974} it has been
widely suspected that there should exist a dual description
of large $N$ gauge theories in terms of string theories.
In these lectures we will discuss a very concrete realization
of such a duality relating type IIB
superstrings in a maximally supersymmetric
plane-wave background to four dimensional
${\cal N}=4$ Super  Yang-Mills Theory in a particular 
double scaling limit, which
was initiated by the influential paper of Berenstein, Maldacena and Nastase
\cite{Berenstein:2002jq} in 2002. 
This  duality may be viewed as a ``corollary'' of the well studied 
Anti-de-Sitter/Conformal Field Theory (AdS/CFT) duality \cite{malda}, 
which asserts a dual description
of the IIB superstring moving in a $AdS_5\times S^5$ 
spacetime-background in terms of the four dimensional 
maximally supersymmetric $U(N)$ gauge theory.
It is probably the
most concrete example of a string/gauge theory duality ever established,
as it is the first to truly probe the ``stringy'' regime of the
correspondence in terms of higher mode 
excitations in the free string theory, as well as higher genus 
worldsheet interactions in the interacting string theory -- a 
regime which has so far been technically
inaccessible in the AdS/CFT correspondence.
In these lecture notes we shall assume a
basic knowledge of string theory, 
gauge theories and the
AdS/CFT duality for which a number of reviews already 
exists \cite{AdSCFT}.

The key developments in establishing the plane-wave string/gauge 
theory correspondence began with the discovery of Blau, Figueroa-O'Farrill, 
Hull and Papadopoulos \cite{Blau:2001ne}, that the type IIB supergravity
solution of a gravitational plane-wave with a constant, null five-form
field strength constitutes a maximally supersymmetric background
for the IIB string. 
As such it is a distinguished background of IIB string theory, 
as there exist only two additional maximally supersymmetric
backgrounds of IIB string theory: The well studied cases
of flat Minkowski and $AdS_5\times S^5$ spaces
\cite{Figueroa-O'Farrill:2002ft}.
In addition it turns out that the plane-wave string reduces to
a free, massive two dimensional model once one goes to the light-cone
gauge, as noticed by Metsaev
and Tseytlin \cite{Metsaev:2001bj,Metsaev:2002re}. It is therefore
as straightforwardly quantized as the superstring in a flat  
background and in this respect strongly distinct to 
the $AdS_5\times S^5$ string, which is given by a
non-linear two dimensional field theory, whose quantization
has not been achieved to date. On the other hand, the plane-wave geometry
is obtained through a limit of the 
$AdS_5\times S^5$ geometry \cite{Blau:2002dy}. 
This is particularly interesting as
by virtue of the AdS/CFT duality this limit must entail
a dual description of the plane-wave string in terms 
of the supersymmetric gauge theory in a corresponding limit 
\cite{Berenstein:2002jq}.
Surprisingly the pane-wave string/gauge theory duality
turns out to be perturbatively
accessible from both sides of the correspondence -- in contradistinction
to the strong/weak coupling duality in AdS/CFT.
It is then possible to set up a concrete ``dictionary''
relating string states to operators in the Super Yang-Mills theory
and to compare their spectra in a perturbative expansion 
on both sides of the correspondence.

The limit to be taken on the gauge theory side is a novel type
of double scaling limit \cite{Berenstein:2002jq},
in which not only the rank $N$ of the
gauge group is taken to infinity, but one is led to simultaneously only
considers correlation functions of operators with a 
diverging R-charge $J\sim \sqrt{N}$. In particular this limit is
of non 't Hooftian type, as the 't Hooft coupling constant
$\lambda:=g_{\rm YM}^2\, N$ diverges. Indeed
the limit maintains contributions from graphs
of all genera \cite{Kristjansen:2002bb,Constable:2002hw},
due to a combinatorial abundance of non-planar 
graphs growing with $J$. Not surprisingly the non-planar sector of the
gauge theory is found to be dual to plane-wave string interactions, which
opens up the possibility of studying string interactions
in the framework of light-cone string field theory via methods of
perturbative large $N$ and $J$ gauge theory. This we shall
do in detail in these lectures. On the string theory side
interactions have been studied
with the methods of light-cone superstring field theory
\cite{Spradlin:2002ar,Spradlin:2002rv,Pankiewicz:2002tg} and
shown to agree with the gauge theory predictions under
certain assumptions, which will
be the subject of the last section.

In summary this novel duality represents itself as an interesting
and very concrete model
to study the complementarity of string and gauge theories.
By doing so one may hope to develop novel and wider accessible
tools which could become useful in the study of 
phenomenologically more interesting systems in the future.

\subsection{Strings and large \mb$N\,$\mn  gauge theories}

The expectation that there should exist a close relationship
between gauge theory and strings is based, among 
other observations, on the analysis 
of the perturbation expansion of a $U(N)$
gauge theory in the large $N$ limit. To understand this in some detail
let us look at the following  schematic action of $N\times N$ hermitian 
matrix fields $(\phi_i)_{ab}(x)$
\be
{\cal S}=\frac{1}{\gym^2}\,\int d^4x\,
 \Bigr [ \Tr(\partial_\mu \phi_i\, \partial^\mu \phi_i)
+c^{ijk}\, \Tr(\phi_i \, \phi_j \, \phi_k) +
d^{ijkl}\, \Tr(\phi_i \, \phi_j \, \phi_k\, \phi_l)\, \Bigr ] \, .
\label{toymodel}
\ee
This action mimics a $U(N)$ Yang-Mills model as well as possible 
couplings of scalar fields
in the adjoint representation. The propagators of the matrix 
valued fields may be represented by ``fat'' graphs
\be
\raisebox{-0.2cm}{\epsfxsize=2.5cm\epsfbox{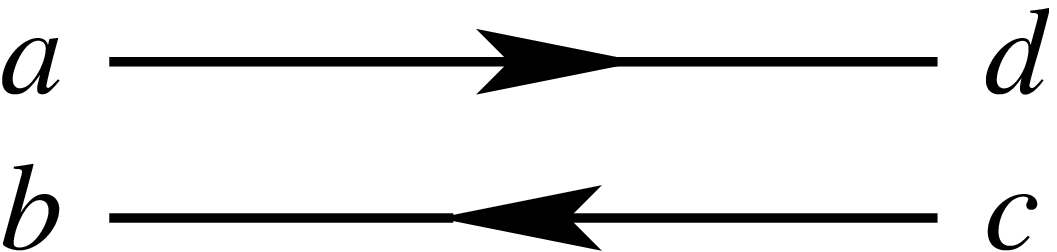}}\quad\sim \gym^2:
\qquad\qquad
\vev{(\phi_i)_{ab}(x)\, (\phi_j)_{cd}(0)}= \frac{\gym^2}{8\pi^2\, x^2}
\, \delta_{ij}\,\delta_{ad}\, \delta_{bc}
\ee
One immediately reads off from
the Lagrangian \eqn{toymodel} that the vertices  scale uniformly with 
$1/\gym^2$. Their fat graph structure may be depicted as follows:
$$
\raisebox{-0.8cm}{\epsfxsize=2cm\epsfbox{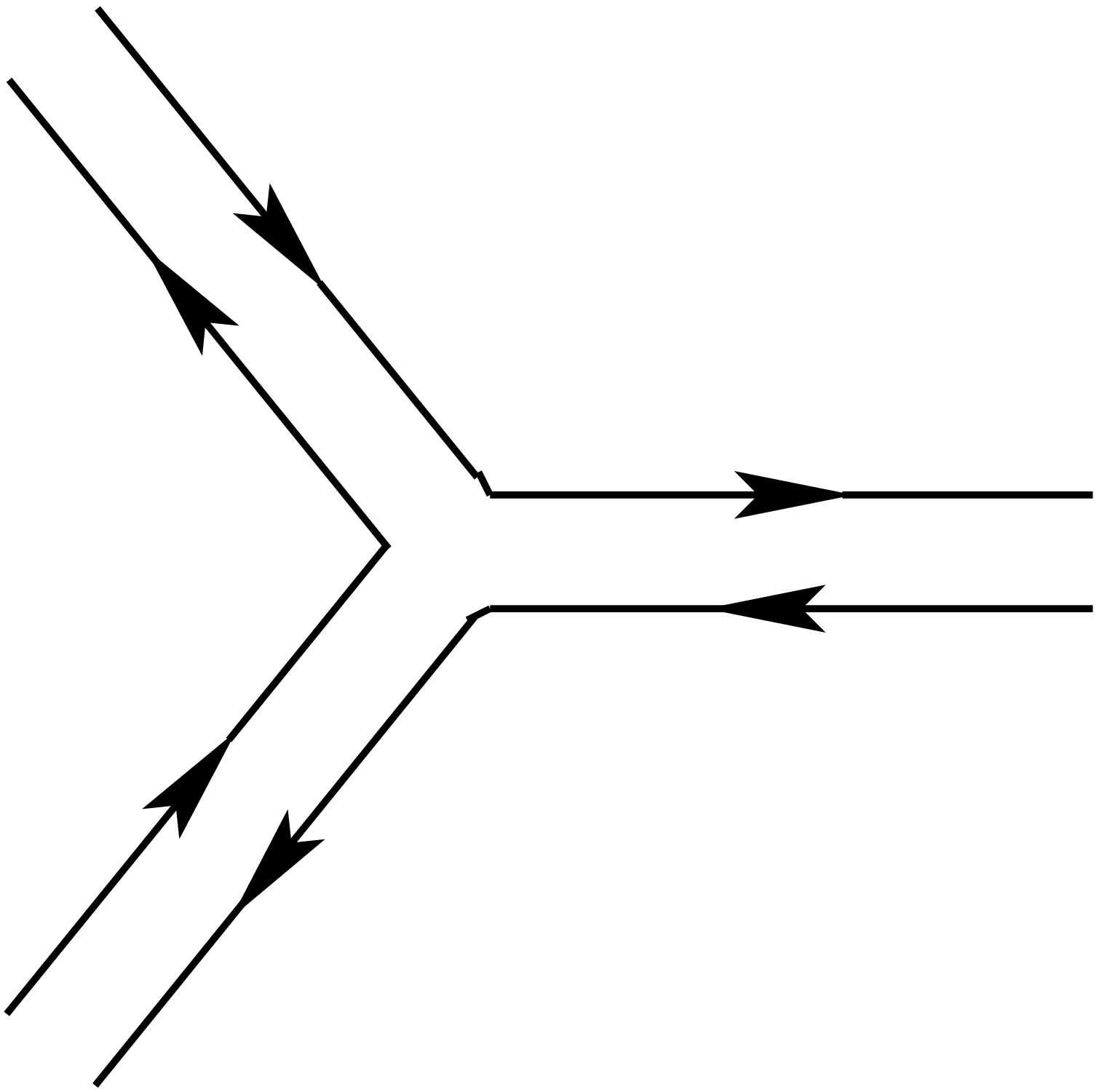}}\quad\sim 
\frac{1}{\gym^2}\qquad\qquad\qquad\qquad
\raisebox{-0.8cm}{\epsfxsize=1.7cm\epsfbox{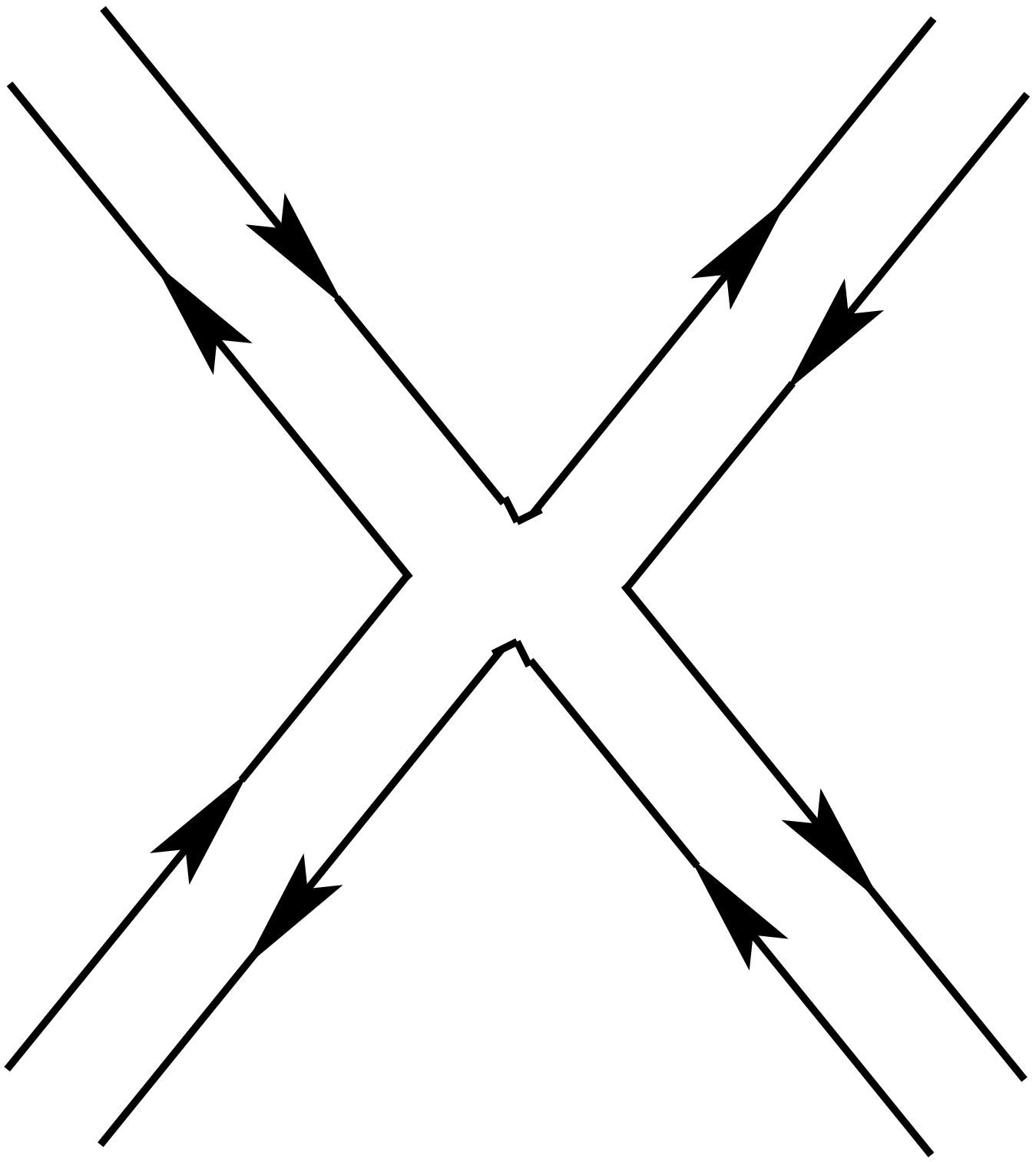}}\quad\sim 
\frac{1}{\gym^2}
$$
By making use of the double line notation for propagators any
Feynman diagram in the perturbative expansion of \eqn{toymodel} 
may be viewed as a simplicial decomposition of a surface with $V$ vertices,
$E$ edges and $F$ faces. Here the total number of propagators in the
Feynman graph corresponds to $E$ and $F$ simply counts the number
of index loops occurring in the graph. As an example let us 
count the factors of $\gym$ and $N$ for the following vacuum 
graphs:
\bea
\raisebox{-0.3cm}{\epsfxsize=0.8cm\epsfbox{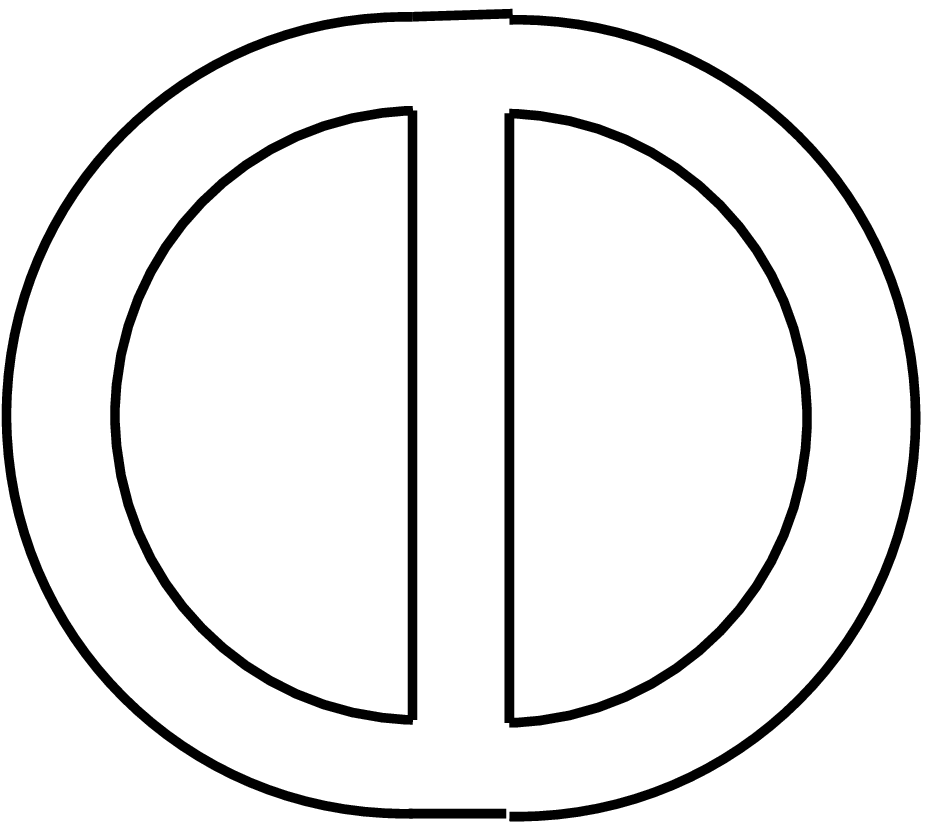}}\,\,
&\entspricht\, (\gym^2)^{3-2}\, N^3 =& (\gym^2\, N)\, N^2\nn\\[0.3cm]
\raisebox{-0.25cm}{\epsfxsize=1cm\epsfbox{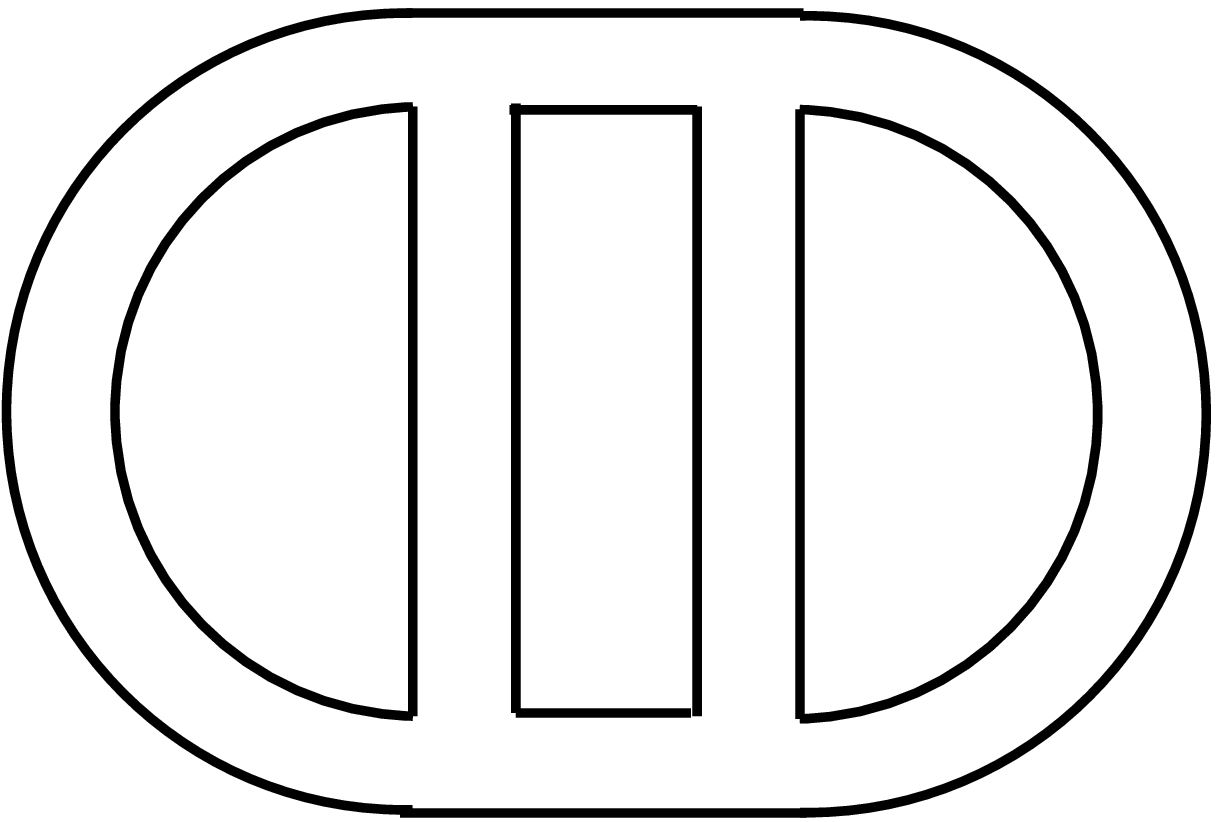}}
&\entspricht\, (\gym^2)^{6-4}\, N^4 =& (\gym^2\, N)^2\, N^2\nn\\[0.3cm]
\raisebox{-0.25cm}{\epsfxsize=1cm\epsfbox{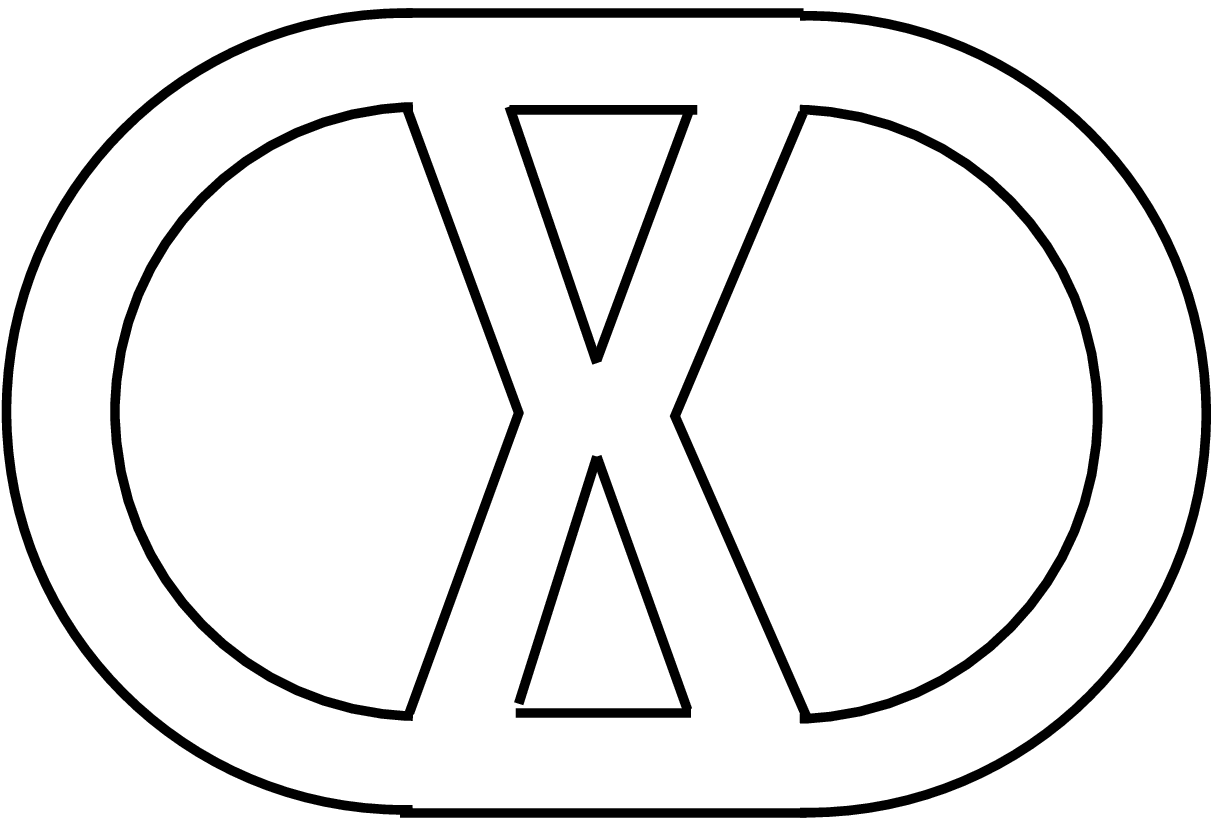}}
&\entspricht\, (\gym^2)^{8-5}\, N^5 =& (\gym^2\, N)^3\, N^2\nn\\[0.3cm]
\raisebox{-0.25cm}{\epsfxsize=1cm\epsfbox{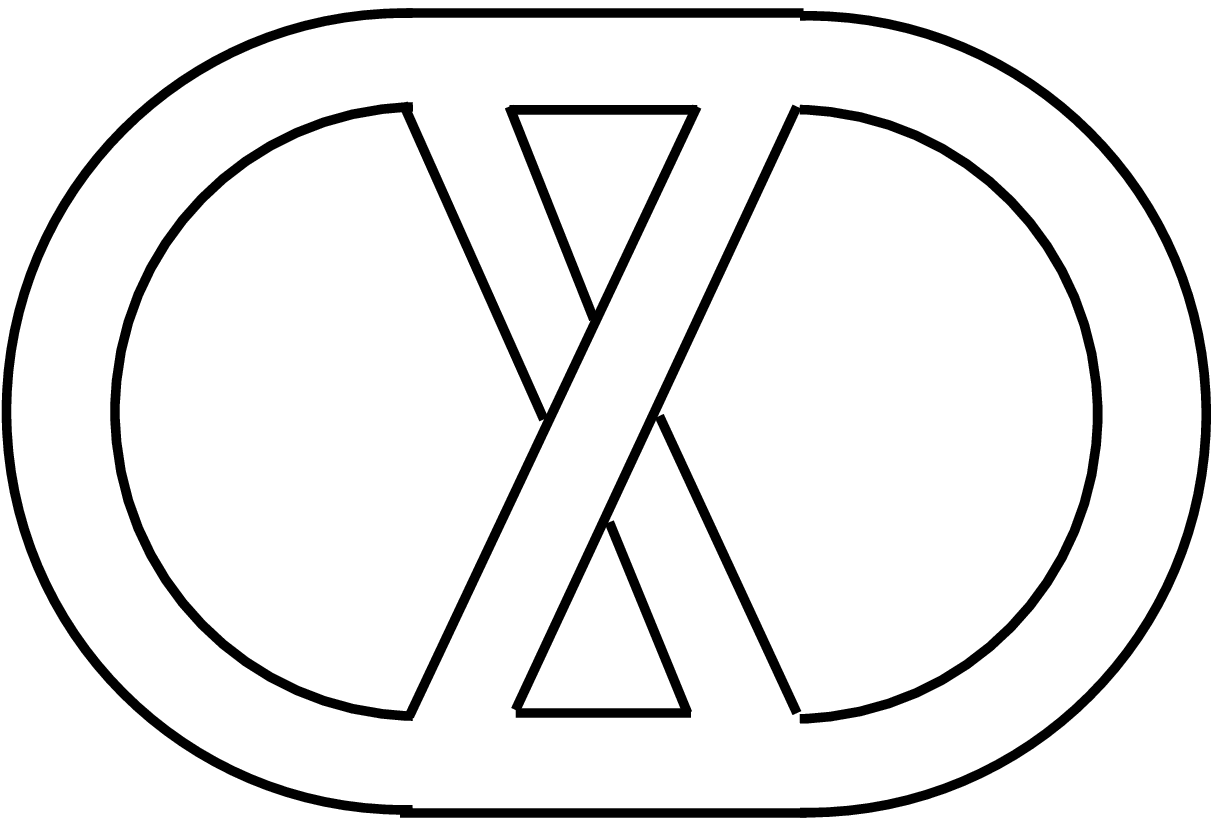}}
&\entspricht\, (\gym^2)^{6-4}\, N^2 =& (\gym^2\, N)^2 
\eea
We observe that the first three graphs are planar, i.e.~they may be
drawn on a plane without crossing of propagators, whereas
the last one is non-planar. Also the combination 
$\lambda=\gym^2\,N$ emerges as a quantum loop counting parameter
known as the 't Hooft coupling constant. Non-planar graphs are
suppressed by powers of $1/N^2$ with respect to planar ones.
In general it is easy to see that for
a graph with $V$ vertices, $E$ propagators and $F$ index loops one has
$$
N^F\, (\gym^2)^{E-V}=N^{V-E+F}\, (\gym^2\, N)^{E-V}
$$
Now the Euler number $\chi$ of a simplicial manifold is given by
$\chi=V-E+F=2-2\,g$, where $g$ denotes the  genus of the manifold,
corresponding to its number of handles. Hence the perturbative 
expansion of \eqn{toymodel}
or a general $U(N)$ gauge theory is organized as a double expansion
in $\lambda:=\gym^2\, N$ and $1/N^2$, counting the number of
quantum loops and handles respectively, i.e.~the free energy $F$
will decompose as
\be
N^2\, F = \sum_{g=0}^\infty N^{2-2g}\, \sum_{n=0}^\infty
c_{g,n}\, \lambda^n
\label{Fexp}
\ee
This implies that there is a consistent way of performing a large
$N$ limit of a gauge theory - due to 't Hooft - by taking $N\to\infty$
while keeping $\lambda$ fixed, i.e.~a simultaneous
scaling of $\gym\sim 1/\sqrt{N}\to0$. Note that
in the strict 't Hooft limit all non-planar graphs are suppressed
and the gauge theory reduces to its planar limit.

The structure of the genus expansion of \eqn{Fexp} strongly 
resembles the perturbative expansion of string theory as a sum over 
worldsheets of growing genus, with the role of $1/N^2$ played
by the string coupling constant $g_s$
$$
{\epsfysize=0.75cm\epsfbox{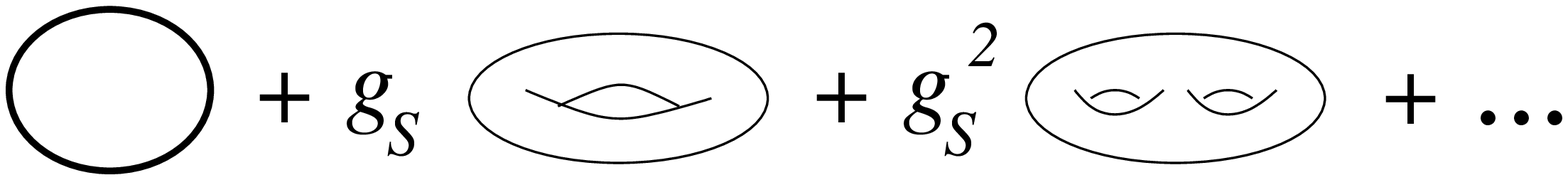}}
$$
As non-planar graphs are suppressed in the strict 't Hooft limit, one
expects that the large $N$ limit of the gauge theory should correspond
to a non-interacting ($g_s=0$) string model. Of course finding the
associated string theory to a given four dimensional gauge theory
has been very hard. 

\subsection{The AdS/CFT Correspondence}

The first concrete proposal
of such a string/gauge theory duality pair emerged more than
20 years after its first suggestion with the AdS/CFT
duality conjecture
due to Maldacena in 1997 \cite{malda}. In its simplest form
it states that the dual string model of the maximally supersymmetric
gauge theory in four dimensions is the type IIB superstring propagating
in the ten dimensional $AdS_5\times S^5$ background geometry. The 
duality conjecture surprisingly relates a four dimensional gauge
theory to a higher dimensional string model, which represents
a manifestation of the so called holographic principle 
\cite{holograpicprinciple} indicating that the entire degrees of freedom of
a quantum theory of gravity reside on the boundary of the space-time
region in question. The boundary of $AdS_5\times S^5$ is four dimensional
and this is where the ${\cal N}=4$ Super Yang-Mills theory lives.
This picture is most transparent in the calculation of Wilson loops
in the dual string model: The contour of the Super Yang-Mills Wilson
loop represents the ends of an open string attached to the boundary of
$AdS_5\times S^5$ extending into the bulk of the anti-de-Sitter space.
The expectation value of the Wilson loop operator is semi-classically
nothing but the minimal surface of the $AdS$-string, which due to 
the curvature extends into the bulk. 

The central relations in the AdS/CFT duality
conjecture relate the gauge theory parameters $\gym$ and $N$ to
the string theory parameters $\alpha'$ (string tension), $g_s$ (string
coupling constant) and
the radius $R$ of the $AdS_5$ and $S^5$ spaces, via
\be
\frac{R^4}{{\alpha}'^2}= \gym^2\, N 
\qquad \mbox{and} \qquad
 4\pi g_s = \gym^2\, .
\ee
Unfortunately though, even the free ($g_s=0$) $AdS_5\times S^5$ string
is a rather complicated two dimensional field theory, whose quantization
remains a very challenging open problem. Hence the string theory side
of the duality conjecture could so far only be addressed by studying its
low energy effective description in terms of type IIB supergravity.
This approximation to string theory is only meaningful as long
as the curvature of the background is small compared to the string
scale, i.e.~the radius $R$ in string units needs to be very large
\be
1\ll \frac{R^4}{{\alpha'}^2}=\lambda
\ee
This domain is perfectly incompatible to the perturbatively accessible
regime of ${\cal N}=4$ Super Yang-Mills, which requires $\lambda\ll 1$!
One is hence dealing with a duality relating a weakly coupled
to a strongly coupled theory and vice versa. So if 
the duality conjecture is indeed true we have a fascinating new
tool at hand for studying the strongly coupled sector of a gauge or string
theory. But in the same instance a proof of the duality conjecture 
is very hard if not impossible, as it requires solving the string 
or gauge theory non-perturbatively. 

As we shall see in the novel plane-wave string/gauge theory duality
to be discussed, 
this situation has improved, as both theories turn out to possess
an overlapping perturbative regime. Therefore the correspondence may
here be tested beyond the supergravity regime into the realm of
true stringy effects. As such the first massive string excitations
may be reproduced in the gauge theory. Moreover the interacting
string sector turns out to correspond to non-planar gauge theory
effects in a novel double-scaling limit. We hence have a very concrete
and testable example of a string/gauge theory duality, which at the least
represents an interesting toy model to hopefully develop new tools for
the study of more complicated and phenomenologically relevant 
string/gauge theory systems.

We should mention that there is a price to pay here. Firstly the
implementation of the
holographic principle in the plane-wave/gauge theory duality is not
well understood at this point\footnote{For work along these lines
see \cite{pwholography,BN}.}. 
The boundary of the plane-wave geometry
is one dimensional, which hints at an
effectively one dimensional dual gauge model - yet undiscovered.
However, we shall encounter first traces of such a reduced 
quantum mechanical model in section 5.
Secondly the new gauge theory limit to be discussed leads us to a vastly
reduced sector of ${\cal N}=4$ Super Yang-Mills: Only a restricted 
class of operators
survive the limit and turn out to directly correspond to the free
string excitations. Moreover only two and three-point functions of
these gauge theory operators turn out to exist in the scaling limit --
higher point functions simply diverge \cite{Beisert:2002bb}! 
Thus the number of possible
observables is strongly reduced. This indeed should be taken as an 
indication of an effective lower dimensional description of the
Berenstein-Maldacena-Nastase (BMN) sector of  ${\cal N}=4$ Super Yang-Mills.

\section{The plane-wave geometry as a Penrose limit of 
\mb$AdS_5\times S^5$\mn}

In this section we shall review the emergence of the plane-wave
geometry as a limit of the $AdS_5\times S^5$ background. The idea
is to zoom into the geometry seen by a particle moving on a
light-like geodesic along a great circle of the $S^5$ sphere 
\cite{Blau:2002dy}. Such a limit is possible for any 
space-time geometry and leads to a plane-wave metric
as pointed out by Penrose \cite{Penrose}.
In global coordinates the $AdS_5\times S^5$ metric is given
by
\be
ds_{AdS_5\times S^5}^2=
R^2\, [\, -dt^2\, \cosh^2\rho + d\rho^2+\sinh^2\rho\, d\Omega_3{}^2
+d\psi^2\, \cos^2\theta + d\theta^2+\sin^2\theta\, d\Omega_3^{\prime\, 2}
\, ] \, .
\label{AdS5xS5}
\ee
One now considers the light-like trajectory parametrized by $\lambda$
along
\be
\rho=0,\qquad \theta=0, \qquad t=t(\lambda),\qquad \psi=\psi(\lambda)
\label{path}
\ee
The relativistic particle moving along this geodesic is governed 
by the action
\be
S=\ft 12 \int d\lambda\, (e^{-1}\, g_{\mu\nu}(x)\, \dot x^\mu\, \dot x^\nu
- e\, m^2) = \ft{R^2}{2}\int d\lambda\, e^{-1}\, (-\dot t^2+\dot \psi^2)
\label{particle}
\ee
with $e$ denoting the ``einbein''. In the second equality
we have specialized to the zero
mass case $m=0$ and the path parametrized in \eqn{path}.
Clearly then upon introducing
light-cone coordinates $\tilde x^\pm=\ft 1 2 (t\pm \psi)$ the light-like
trajectory given by $\tilde x^-=\lambda$ and $\tilde x^+=\mbox{const.}$
solves the equations of motion arising from \eqn{particle}.
In order to study the geometry near this trajectory we introduce
the new coordinates $x^\pm,r,y$ in the particular
scaling limit $R\to\infty$
\be
x^+=\frac{\tilde x^+}{\mu}, \qquad x^-=\mu\, R^2\, \tilde x^-, \qquad
\rho=\frac{r}{R},\qquad \theta=\frac{y}{R}
\label{newvars}
\ee
where $\mu$ is a new mass parameter introduced in order to maintain
canonical length dimensions for $x^\pm,r$ and $y$. If one now performs this
change of variables in \eqn{AdS5xS5} one sees that the terms of order
$R^2$ cancel out and the leading contributions are $R$ independent
\bea
ds_{AdS_5\times S^5}^2 &=& R^2\, [ -\mu^2\, (dx^+)^2 +\mu^2\, (dx^+)^2 ]
+ [ -2\, dx^+\, dx^- - \mu^2\, r^2\, (dx^+)^2 + dr^2+r^2\, d\Omega_3{}^2
\nn\\&&
-2\, dx^+\, dx^- - \mu^2\, y^2\, (dx^+)^2 + dy^2+y^2\, d\Omega_3^{\prime\, 2}
] + {\cal O}(R^{-2})\nn\\
&=&-4\, dx^+\, dx^- - \mu^2\, 
(\vec y^2 +\vec r^2)\, (dx^+)^2 + d\vec y^2 + d\vec r^2 + {\cal O}(R^{-2})
\label{last}
\eea
where we have introduced the four-vectors $\vec r$ and $\vec y$ in the
last step. Hence in the Penrose limit $R\to \infty$ the 
$AdS_5\times S^5$ metric approaches the plane-wave metric
\be
ds_{AdS_5\times S^5}^2 \to ds_{\rm pw}^2 = -4\, dx^+\, dx^- - \mu^2\, 
(x^i)^2\, (dx^+)^2 + (dx^i)^2\qquad i=1,\ldots,8
\label{pwmetric}
\ee
Similar considerations may be applied to the non-vanishing self-dual
five-form to yield 
\be
F_{+1234}=F_{+5678}=4\mu
\label{5form}
\ee
in the Penrose limit. Therefore the transverse $SO(8)$ invariance
of the metric \eqn{pwmetric} is broken to a $SO(4)\times SO(4)$
subgroup by the five-form field strength. In the light-cone string
action this breaking will manifest itself in the fermionic mass term.
We also observe that by taking the mass parameter $\mu$ to zero
the plane-wave geometry contracts to flat Minkowski space-time.
Hence, at least on the string theory side all results should
limit to the well known flat background scenario upon taking
$\mu$ to zero. In the dual gauge theory we shall see that
$\mu\to 0$ corresponds to the strict strong coupling limit.

How does the Penrose limit, $R\to \infty$, translate into the
dual gauge theory? For this it is instructive to study how the
energy $E=i\partial_t$ and angular momentum $J=-i\partial_\psi$
conjugate to the global coordinates $t$ and $\psi$ relate to 
the newly introduced light-cone quantities $x^\pm$ and their
conjugate momenta,
\bea
{\cal H}_{\rm lc}:= 
2\, p^- &=& i\partial_{x^+} = \mu\, i (\partial_t+\partial_\psi)=
\mu\, (E-J)\nn\\
2\, p^+ &=& i\partial_{x^-} = \frac{1}{\mu\, R^2}
 i (\partial_t-\partial_\psi)= \frac{E+J}{\mu\, R^2}\, ,
\label{lcscaling}
\eea
where we identified the light-cone Hamiltonian ${\cal H}_{\rm lc}$
with $2\, p^-$. In the limit $R\to \infty$ we see that generic
excitations (corresponding to string states in this background) 
will have vanishing $p^+$ momenta, unless the angular 
momentum $J$ of such a state grows with $R$ as $J\sim R^2$ in a correlated
manner. 
In order to maintain a finite light-cone momentum for such a state, one
deduces the further requirement from \eqn{lcscaling} that
 $E\approx J$ in the Penrose limit\footnote{Note that $p^\pm$ are 
non-negative due to the BPS condition $E\geq |J|$.}. As was pointed out in
the initial paper of Berenstein, Maldacena and Nastase 
\cite{Berenstein:2002jq} the standard AdS/CFT correspondence
linking ${\cal N}=4$ Super Yang-Mills to type IIB strings in
$AdS_5\times S^5$ must entail a new duality 
relating plane-wave strings to an adequate limit of the ${\cal N}=4$
gauge theory, corresponding to the discussed 
Penrose limit. The nature of this limit
may be identified by translating the gravity quantities $E$ and
$J$ to Super Yang-Mills variables. Here the energy $E$ in global
coordinates is identified with the scaling dimension $\Delta$ of
a composite Super Yang-Mills operator. The angular momentum $J$ on the other
hand corresponds to the charge of a $U(1)$ subgroup of the $SO(6)$
$R$ symmetry group of ${\cal N}=4$ Super Yang-Mills, which we shall
discuss in more detail later on. Therefore the first relation
of \eqn{lcscaling} may be rephrased from the gauge theory perspective
as 
\be
\frac{{\cal H}_{\rm lc}}{\mu}\entspricht \Delta -J
\label{BMNrelation}
\ee
which is the central relation in the BMN correspondence. Due to the
AdS/CFT relation $R^4={\alpha'}^2\, \gym^2\, N$ the Penrose limit
$R\to\infty$ with $J\sim R^2$ translates into the gauge theory limit
\be
N\to \infty, \qquad J\sim \sqrt{N},\qquad \gym\,\,\mbox{held fixed}
\label{BMNlimit1}
\ee
Holding $\gym$ fixed in this limit corresponds to a finite
value of the string coupling constant $g_s=\gym^2/4\pi$ on
the dual string side.
Moreover the finite light-cone
energy requirement $E\approx J$  discussed above tells us that in the 
gauge theory limit only Super Yang-Mills operators with
\be
\Delta\approx J
\label{EsimJ}
\ee
will survive and correspond to finite light-cone energy states
on the string side. We will return to this gauge theory limit in
section four.

\section{Light-cone quantization of the type IIB plane-wave string}

Let us now discuss the quantization of the
type IIB superstring in the plane-wave background 
\eqn{pwmetric} and \eqn{5form}. Due to the non-vanishing
Ramond-Ramond background field strength it is
necessary to work with the Green-Schwarz formulation of the
superstring, defined through the worldsheet fields $X^\mu(\tau,\sigma)$
and $\theta^A_\alpha(\tau,\sigma)$ with $\mu=0,\ldots,9$, $A=1,2$ and 
$\alpha=1,\ldots,16$ being 10d space-time vectors and two Majorana-Weyl
spinors of same chirality respectively\footnote{For an introduction
to superstrings in the Green-Schwarz formulation see chapter five
of \cite{GSW}.}. 
The resulting covariant action in the plane-wave 
background was worked out by Metsaev \cite{Metsaev:2001bj} and
takes a very complicated form consisting of terms up to order
${\cal O}(\theta^{16})$ in the fermionic sector.
The model becomes tractable, however, in the light-cone gauge
\cite{Metsaev:2001bj,Metsaev:2002re}. For this one uses the 2d 
diffeomorphism invariance to go to the conformal gauge for the
worldsheet metric 
$
g_{ab}=e^{\phi}\, \left ( \begin{matrix} -1&0\cr 0& 1\cr
\end{matrix}\right )
$.
The freedom of performing residual conformal transformations
is then used to set 
\be
X^+(\tau,\sigma)= p^+\, \tau
\label{lcg}
\ee
in complete analogy to the light-cone quantization in 
flat Minkowski background. In the fermion sector 
the local fermionic
$\kappa$-symmetry is employed to gauge away one half of the fermionic
degrees of freedom via the condition $\Gamma^+\, \theta^A=0$, again
as is done in the flat background \cite{GSW}. This
gauge choice dramatically simplifies the fermionic
sector of the model and only terms up to quadratic order in fermions
survive\footnote{The emergence of a quadratic string action 
in the Penrose limit may also 
be traced back to the semi-classical quantization of the full 
$AdS_5\times S^5$ superstring action around a point-like solution
of the classical string equations of motion propagating on a geodesic
along the $S^5$, see \cite{semicl1,semicl2} for details.}. 
Explicitly one obtains a free quadratic model with
action
\bea
S&=&\frac{1}{2\pi\alpha'}\int_{-\infty}^{\infty}d\tau \int_0^{2\pi\alpha'p^+}
d\sigma\, \Bigl( \ft 1 2 \,\partial_a X^I\,\partial_a X^I - \ft 1 2\, \mu^2
(X^I)^2 \nn\\
&& \qquad\qquad+ i\, \theta^1(\partial_\tau+\partial_\sigma)\theta^1
+i\, \theta^2(\partial_\tau-\partial_\sigma)\theta^2 -2\mu\, \theta^1
\Gamma_{1234}\theta^2\Bigr)\, .
\label{ppwavestring}
\eea
The bosonic mass term stems from the worldsheet
coupling of the background space-time metric in the $++$ direction, i.e.
\be
g_{++}\, \partial_a X^+\partial_a X^+=-\mu^2\, (X^I)^2\, ,
\ee
due to the light-cone-gauge condition \eqn{lcg}.
Similarly  the fermionic mass term arises from the fermion bilinear
coupling to the Ramond-Ramond field strength $F_{+1234}$, which manifestly
breaks the transverse $SO(8)$ to $SO(4)\times SO(4)$. Also we see
that in the limit $\mu\to 0$ one recovers the standard flat space model
\cite{GSW}. 

The bosonic equations of motion following from \eqn{ppwavestring}
take the form
\be
(\partial_\tau^2-\partial_\sigma^2 +\mu^2)\, X^I=0
\ee
subject to the closed string
boundary condition $X^I(\tau,\sigma+2\pi\, \alpha'\, p^+)=
X^I(\tau,\sigma)$. Its general solution in an oscillator 
mode decompositions
reads
\be
X^I=\cos (\mu\tau)\, \frac{x_0^I}{\mu} + \sin(\mu\tau)\, \frac{p_0^I}{\mu} +
\sum_{n\neq 0}\frac{i}{\sqrt{2\,\omega_n}}\,(\,
 \alpha^I_n\, e^{-i(\omega_n\tau-k_n\sigma)}
+ \tilde\alpha^I_n\, e^{-i(\omega_n\tau+k_n\sigma)}\, ) \, ,
\ee
where $\omega_n=\mbox{sign}(n)\, \sqrt{k^2_n+\mu^2}$ and
$k_n=n/(\alpha'\, p^+)$. The corresponding canonical momentum,
$P^I=\dot X^I$, reads
\be
P^I=\cos (\mu\tau)\, p_0^I - \sin(\mu\tau)\, x_0^I +
\sum_{n\neq 0}\sqrt{\frac{\omega_n}{2}}\,(\,
 \alpha^I_n\, e^{-i(\omega_n\tau-k_n\sigma)}
+ \tilde\alpha^I_n\, e^{-i(\omega_n\tau+k_n\sigma)}\, ) \, .
\ee
Moreover the coordinate $X^-$ is expressed in terms of the transverse
degrees of freedom via the Virasoro constraint
\be
P^+\,\partial_\sigma X^- + P^I\, \partial_\sigma X^I+
i \theta^1\,\partial_\sigma\theta^1 + i \theta^2\,\partial_\sigma\theta^2
=0\,  ,
\ee
which arises as a consequence of the conformal gauge choice. Next to
determining $X^-(\tau,\sigma)$, the above equation also implies 
a constraint on the transverse degrees of freedom upon integrating it
over $\sigma$:\footnote{Note that $P^+$ is a constant due to the light-cone
gauge condition \eqn{lcg}. Also we demand 
$X^-(\tau,\sigma+2\pi\alpha'p^+)=X^-(\tau,\sigma)$.}
\be
\int_0^{2\pi\alpha'p^+}d\sigma\, [\,  P^I\,\partial_\sigma X^I +
i \theta^1\,\partial_\sigma\theta^1 + i \theta^2\,\partial_\sigma\theta^2\,
 ] =0\, .
\label{vircstr}
\ee
The plane-wave string is now readily quantized by replacing
Poisson brackets by commutators in the standard fashion
$
\{ . , .\}_{\rm P.B.} \rightarrow i\, [ . , .]
$.
From the canonical commutation relations 
one then deduces the commutation relations for the modes
\be
[p^I_0,x^I_0]= -i \delta^{IJ},\qquad [\alpha^I_m,\tilde\alpha^J_n]=0, \qquad
[\alpha^I_m, \alpha^J_n] = \delta_{n+m,0}\, \delta^{IJ},\qquad
[\tilde\alpha^I_m, \tilde\alpha^J_n] = \delta_{n+m,0}\, 
\delta^{IJ} \, .
\ee
Similar expressions arise for the fermionic modes, for details see
\cite{Metsaev:2002re}. The Hamiltonian then takes the form
\be
{\cal H}_{\rm lc}= \frac{1}{p^+}\, \int d\sigma\, \Bigl [
\, (P^I)^2 +(\partial_\sigma X^I)^2 + \mu^2\, (X^I)^2 + \mbox{fermions}
\, \Bigr ] \, .
\label{SH01}
\ee
It is useful to introduce modes in the zero mode sector of
$x_0^I$ and $p_0^I$ as well via
\be
\alpha_0^I=\frac{1}{\sqrt{2\,\mu}}\, (\,p_0^I+i\,\mu\, x_0^I\, ),
\ee
in order to write down the Hamiltonian \eqn{SH01} in terms of oscillator
modes 
\be
{\cal H}_{\rm lc} = \mu \, (\alpha_0^{\dagger\, I}\, \alpha_0^I
+ \theta_0^\dagger\, \theta_0\, )
+ \frac{1}{\alpha'\, p^+}\, \sum_{n=1}^\infty \sqrt{n^2 + (\alpha'\, p^+\,
\mu)^2}\, \Bigl [ \, \alpha_{-n}^I\, \alpha_n^I + \tilde\alpha_{-n}^I\, 
\tilde\alpha_n^I + \theta^1_{-n}\, \theta^1_n+ \theta^2_{-n}\, \theta^2_n\,
\Bigr ]
\label{HLK1}
\ee
where we have now also included the fermion modes in the conventions
of \cite{Metsaev:2002re}.
An immediate observation is that upon taking $\mu\to 0$ the flat
space light-cone Hamiltonian arises. This is indeed necessary, as
for $\mu=0$ the plane-wave geometry \eqn{pwmetric} turns into 
flat Minkowski space. The most pronounced difference of this 
Hamiltonian compared to
the flat space situation is the fact that also
the zero-mode sector is massive and governed by a harmonic oscillator
spectrum. Therefore there are no asymptotically free states in the
transverse direction anymore, rendering the concept of an S-matrix
for the plane-wave string problematic: {\sl All} transverse
excitations are bound in a harmonic well.

One furthermore defines the Fock-vacuum $|0,p^+\rangle$ to be
annihilated by the positive modes
\bea
\alpha_0\, |0,p^+\rangle=0, & \alpha_n\, |0,p^+\rangle=0, & 
\tilde\alpha_n\, |0,p^+\rangle=0, \qquad n\geq 1 \nn\\
\theta_0\,  |0,p^+\rangle=0, & \theta_n^1\,  |0,p^+\rangle=0, & 
\theta_n^2\,  |0,p^+\rangle=0, 
\eea
The physical states are subject to the Virasoro constraint arising
from \eqn{vircstr}
\bea
(N-\tilde N)\, |{\rm phys}\rangle=0 \qquad
{\rm with}\quad 
 N&=&\sum_{n=1}^\infty (\, \alpha_{-n}^I\,\alpha_n^I +\theta^1_{-n}\,
\theta^1_n\, )\nn\\
{\rm and}\quad 
 \tilde N&=&\sum_{n=1}^\infty (\, \tilde\alpha_{-n}^I\,
\tilde\alpha_n^I +\theta^2_{-n}\,
\theta^2_n\, )
\label{levelmatching}
\eea
requiring a balanced excitation structure of the two sets of modes.
The resulting spectrum then takes the simple form
\be
E_{\rm lc}= \mu\, N_0 + \mu\, (\, N_n + \tilde N_n\, ) \, 
\sqrt{1+\frac{n^2}{(\alpha'\, p^+\, \mu)^2}} \, .
\label{spectrum}
\ee
Let us now write down the lightest bosonic excitations. In the zero-mode
or supergravity sector one has
\bea
|0,p^+\rangle && E_{\rm lc}=0 \nn\\
\alpha_0^{\dagger \, I}\, |0,p^+\rangle && E_{\rm lc}=\mu \nn\\
\theta_0^{\dagger}\, |0,p^+\rangle && E_{\rm lc}=\mu \nn\\
\alpha_0^{\dagger\, I_1}\ldots
\alpha_0^{\dagger\, I_N}\, |0,p^+\rangle && E_{\rm lc}=N\cdot \mu 
\label{sugraspectrum}
\eea
and the first true stringy excitations - paying attention to the
level matching condition \eqn{levelmatching} - read
\bea
\alpha_{-n}^I\, \tilde\alpha_{-n}^J|0,p^+\rangle && E_{\rm lc}=2
\, \mu\, \sqrt{{1+\frac{n^2}{(\alpha'\, p^+\, \mu)^2}}} \nn\\
\alpha_{-n_1}^I\,\alpha_{-n_2}^J\, \tilde\alpha_{-n_3}^K
|0,p^+\rangle && E_{\rm lc}=
\sum_{i=1}^3
\mu\, \sqrt{{1+\frac{n_i^2}{(\alpha'\, p^+\, \mu)^2}}}
\qquad \mbox{with}\quad n_1+n_2=n_3\nn\\
&\vdots& 
\label{stringyspectrum}
\eea
Hence the spectrum of free, non-interacting, plane-wave string theory
is under complete control. In the following sections 4 and 5 we shall see
how it is reproduced from the dual gauge theory in terms of the
scaling dimensions of the associated Super Yang-Mills operators.

String interactions may be described by methods of light-cone
string field theory in the plane-wave background and will be
the subject of section 6.

\section{Plane-wave strings from \mb${\cal N}=4\,\, $\mn Super Yang-Mills}

As was pointed out in the initial paper of Berenstein, Maldacena
and Nastase \cite{Berenstein:2002jq} the discussed Penrose limit
of $AdS_5\times S^5$ leading to the plane-wave geometry must
entail -- by virtue of the AdS/CFT correspondence -- 
a dual gauge theory description of the plane-wave string model. 
In this section we shall determine the precise nature of this 
dual gauge theory limit.

The dual gauge theory of the $AdS_5\times S^5$ superstring is
the maximally supersymmetric (${\cal N}=4$)
Yang-Mills theory in four dimensions \cite{N4SYM}.
Its field content is comprised of a gluon field, six scalars as well as 
4 Majorana gluinos, which we choose to write as a 16 component
10d Majorana-Weyl spinor. All fields are in the adjoint representation 
of $U(N)$. Explicitly
we have
\be
A_\mu(x), \qquad
\phi_i(x),\quad i=1,\ldots, 6 \qquad
\chi_\alpha(x), \quad \alpha=1,\ldots 16
\label{fields}
\ee
given by $N\times N$ hermitian matrices. The action of ${\cal N}=4$
Super Yang-Mills is uniquely determined by two parameters, the
coupling constant $g_{\rm YM}$ 
and the rank of the gauge group $N$, to be
\be
S=\frac{2}{\gym^2}\, \int d^4x\, \Tr\, \Bigl\{ \frac{1}{4}\, (F_{\mu\nu})^2
+\frac{1}{2}\, (D_\mu\phi_i)^2 -\frac{1}{4}\, [\phi_i,\phi_j]\,
[\phi_i,\phi_j] + \frac{1}{2}\, \bar\chi D\!\!\!\!/\,\, \chi - \frac{i}{2}
\, \bar \chi\, \Gamma_i\, [\phi_i,\chi]\, \Bigr \}
\label{N4SYM}
\ee
with the covariant derivative defined as $D_\mu= \partial_\mu
-i[A_\mu,\,\,\,]$. Furthermore, $(\Gamma_\mu,\Gamma_i)$ are the
ten dimensional Dirac matrices.

This model displays a global $SO(6)$ symmetry group, called R-symmetry,
acting as internal rotations on the six scalars and four spinors. 
Moreover due to the
large amount of supersymmetry present, the conformal invariance 
of the classical field theory survives the quantization procedure: The
coupling constant $\gym$ is not renormalized and its $\beta$-function
is believed to vanish to all orders in perturbation theory
\cite{N4SYMconf}.
This is why one often refers to ${\cal N}=4$ Super Yang-Mills
as a ``finite'' quantum field theory. The conformal invariance 
group in four dimensions consisting of the Poincare group,
dilatations and special conformal transformations is
$SO(2,4)$.
The full bosonic symmetry group of \eqn{N4SYM} is hence given by
the product $SO(2,4)\times SO(6)_R$ -- matching precisely with
the isometry groups of the $AdS_5\times S^5$ geometry. 

The vertices and propagators can be read off from the action
\eqn{N4SYM}.
We will work in the Feynman gauge, where the free field limit of
the gluon propagators in matrix notation is
\be
\langle \, (A_\mu)_{ab}(x)\,  (A_\nu)_{cd}(y)\, \rangle_0
=\frac{\gym^2\, \delta_{\mu\nu}}{8\, 
\pi^2\, (x-y)^2}\, \delta_{ad}\, \delta_{bc}
\, ,
\label{prop1}
\ee
similarly the propagators for the scalars read
\be
\langle \, (\phi_i)_{ab}(x)\,  (\phi_j)_{cd}(y)\, \rangle_0
=\frac{\gym^2\, \delta^{ij}}{8\, 
\pi^2\, (x-y)^2}\, \delta_{ad}\, \delta_{bc}
\, .
\label{prop2}
\ee

The observables of interest to us are local, composite, 
gauge invariant operators, i.e.~traces of products of fundamental 
fields at a given space-point, 
e.g.~${\cal O}_{i_1\ldots i_k}(x)=\Tr[\phi_{i_1}(x)\,\phi_{i_2}(x)\ldots 
\phi_{i_k}(x)]$. A central class of operators in a general conformal
field theory are the conformal primary operators, which possess a definite
scaling dimension. Their two point functions are determined by the
conformal symmetry to be diagonal and to take the form
\be
\langle {\cal O}_A(x)\, {\cal O}_B(y)\rangle = \frac{\delta_{AB}}{(x-y)^{2\,
\Delta_{{\cal O}_A}}}
\label{CFT2PtFct}
\ee
where $\Delta_{{\cal O}_A}$ is the  scaling dimension of the composite
operator ${\cal O}_A$. Classically these scaling dimensions are simply
the sum of the individual dimensions 
of the constituent fields ($[\phi_i]=[A_\mu]=1$ and $[\chi]=3/2$). 
In quantum theory the scaling dimensions
receive radiative correction, organized in
a double expansion in $\lambda=\gym^2\, N$ (loops) and $1/N^2$ (genera)
\be
\Delta = \Delta_0 + \sum_{l=1}^\infty \lambda^l\, \sum_{g=0}^\infty
\frac{1}{N^{2\, g}}\, \Delta_{l,g} \, ,
\label{anexp}
\ee
as discussed in section one.
For example the simplest conformal primary operator of ${\cal N}=4$
Super Yang-Mills is the Konishi field ${\cal O}_K=\Tr [\phi_i\, \phi_i]$,
whose planar scaling dimension is known up to two loop order 
$\Delta_{{\cal O}_K}=2 + \frac{3\, \lambda}{4\pi^2}
- \frac{3\, \lambda^2}{16\pi^4}$
\cite{Bianchi:2000hn,Arutyunov:2001mh}\footnote{The three loop
result has been recently conjectured to be $\frac{21\, \lambda^3}{256\pi^6}$
\cite{Beisert:2003tq}. Strictly speaking the 't Hooft expansion of
anomalous dimensions \eqn{anexp} can also contain odd powers of $1/N$
due to operator mixing effects, see e.g.~\cite{Beisert:2003tq} for a discussion.}. 

A remarkable feature of the ${\cal N}=4$ gauge theory is the
existence of a class of operators, referred to as 
chiral primary or 1/2 BPS operators, whose scaling dimensions 
do {\sl not} receive any radiative corrections. They can be conformal
primaries or descendants thereof. In the scalar 
sector these protected operators are given by 
\be
{\cal O}^k_{\rm CPO}= C_{i_1i_2\ldots i_k}\, \Tr[\phi_{i_1}\, \phi_{i_2}
\ldots \phi_{i_k}]
\label{CPO}
\ee
with $C_{i_1i_2\ldots i_k}$ being a symmetric traceless rank $k$
tensor. The claim then is that the {\sl exact} scaling dimension 
of ${\cal O}^k_{\rm CPO}$ is
given by its classical value $\Delta_{{\cal O}^k_{\rm CPO}}=k$.

Conformal symmetry moreover  constrains the three-point functions
of conformal primary operators. Their space-time dependence
is completely determined by the scaling dimensions $\Delta_i$
of the participating operators
\be
\langle\, {\cal O}_1(x_1)\,{\cal O}_2(x_2)\,{\cal O}_3(x_3)\,\rangle
=\frac{C_{123}}{|x_1-x_2|^{\Delta_1+\Delta_2-\Delta_3}\,
|x_2-x_3|^{\Delta_2+\Delta_3-\Delta_1}\,
|x_3-x_1|^{\Delta_3+\Delta_1-\Delta_2}}
\ee
and the only new quantity emerging is the structure constant $C_{123}$.
Three-point functions of protected operators are also protected
from radiative corrections, four and higher-point functions
do receive radiative corrections.

In order to address the BMN limit of the ${\cal N}=4$ model
we need to identify the $U(1)$ charge $J$ in the gauge theory
corresponding to the angular momentum $J$ along the 
equator of the $S^5$ on the
dual string side. It is given by the charge associated to the
complex combination of two scalars, say $\phi_5$ and $\phi_6$
\be
Z=\frac{1}{\sqrt{2}}\, (\phi_5+ i\,\phi_6) \, .
\ee
The classical scaling dimensions and $J$ charges of the fundamental
fields are summarized in table \ref{figure1}. Note that the 
fermions $\chi_\alpha$
split into two components $\psi_A$ and $\tilde\psi_{\dot A}$ of opposite
$J$ charge \cite{Berenstein:2002jq}. 
\begin{table}[t]
\begin{center}
\begin{tabular}{|c|rrcr|cc|}
\hline 
 & $Z$ & $\bar Z$ & $\phi_{i=1,2,3, 4}$ & $A_\mu$ & $\psi_A$ & 
$\tilde\psi_{\dot A}$ \cr
\hline 
$\Delta_0$& 1 & 1 &1 &1 & $3/2$ & $3/2$ \cr
$J$ & 1 & -1 & 0 & 0 & $1/2$ & $-1/2$\cr
\hline
$\Delta_0-J$ & 0 & 2 & 1 & 1 &1 & 2 \cr \hline
\end{tabular}
\end{center}
\caption{Scaling dimensions and $J$ charges of the fundamental fields.}
\label{figure1}
\end{table}

As discussed in section 2 the limit in question is then
\be
N\to \infty \quad \mbox{and}\quad J\to \infty \quad \mbox{with}
\quad \frac{J^2}{N}\quad \mbox{and}\quad \gym \quad \mbox{fixed}
\label{BMNlimit}
\ee
Let us stress that this limit is {\sl distinct} to the standard
't Hooft large-$N$ limit, where one takes $N\to\infty$ while
keeping $\lambda=\gym^2\, N$ fixed. In the limit of \eqn{BMNlimit}
$\lambda$ diverges, which seems 
disastrous from a perturbative point of view, as the quantum loop  
corrections in the gauge theory appear as an expansion in $\lambda$.
However, there is one central point we have not yet addressed, which 
was raised in the discussion at the end of section two, namely the necessity 
of restricting one's attention
to the set of operators whose scaling dimensions are of the order
of $J$, i.e.~$\Delta \sim J$. Therefore the operators obeying this
rule are made out of a long string of $Z$'s, compare table \ref{figure1}.

For the class of protected operators \eqn{CPO} the strong coupling 
nature of the BMN limit \eqn{BMNlimit} is not visible
as their two and three-point functions do not receive any
quantum corrections. Examples obeying $\Delta\sim J$ are
\be
\Tr(Z^J) \quad  \mbox{and} \quad \Tr(\phi_i\, Z^J)\, .
\ee
But one can do slightly better than that: The crucial insight of 
Berenstein, Maldacena and Nastase was to violate
the ``protectedness'' of these operators in a small and controlled fashion,
by inserting in the string of $J$ $Z$'s a small number of impurities in
form of operators with $\Delta-J=1$
$$
\Tr(\, \phi_i\, Z\ldots Z\, \phi_j\, Z\ldots Z\, D_\mu Z\, Z\ldots Z\, 
\psi_\alpha\, Z\ldots Z\, ) \, ,
$$
for a generic ``BMN-operator''\footnote{This is actually not
entirely correct: $SO(4)$ singlet operators also require the
compensating insertion of $\bar Z$'s with $\Delta-J=2$ 
\cite{Beisert:2002bb}.}.
As we will show in the following two and three-point functions of 
these type of operators 
receive quantum correction through an {\sl effective} loop
counting parameter
\be
\lambda':=\frac{\gym^2\, N}{J^2}
\ee
which remains finite and tunable in the BMN limit \eqn{BMNlimit}. 
Hence, even though 
the scaling dimensions of generic operators in the ${\cal N}=4$ model
diverge in 
the limit \eqn{BMNlimit}, there remains a perturbatively accessible
sector comprised of BMN-operators with dimensions expressed in terms
of the  new effective coupling constant $\lambda'$. 

Notably this effective weak coupling sector breaks down once one moves
to four and higher point functions. As was demonstrated in 
\cite{Beisert:2002bb} four-point functions of  
$\Tr(Z^J)$ diverge with
$J$ in the BMN-limit. This shows that the BMN limit \eqn{BMNlimit}
represents an extreme reduction of the quantum field theory 
whose precise nature remains to be understood. 

\subsection{The plane-wave string state/gauge theory operator dictionary}

Can we identify the gauge theory operators which are dual to the
plane-wave string states constructed in section 3? The guiding principle
is the value of $\Delta-J$ to be identified with $E_{\rm lc}/\mu$ of
\eqn{spectrum} by the central relation \eqn{BMNrelation}.
For the string groundstate $|0,p^+\rangle$
with vanishing groundstate energy there is a unique
single trace operator with vanishing $\Delta-J$, i.e.
\be
E_{\rm lc}=0\qquad
|0,p^+\rangle \,\,\entspricht \,\,\ft{1}{\sqrt{J\, N^J}}\, \Tr\, Z^J
\qquad \Delta-J=0
\ee
As $\Tr Z^J$ is a protected operator, its scaling dimension equals
$J$ to all orders in $\lambda'$ in the full quantum theory. 
This is indeed necessary for
the above identification to make sense. The normalization of 
$\Tr Z^J$ is chosen in order to have unit weight in the two point
function \eqn{CFT2PtFct} at leading order in $N$.

Let us now move on to the supergravity modes in the plane-wave string 
spectrum \eqn{sugraspectrum}. Here we need $\Delta-J=1$ for the first
excitations. This is realized by the operators
\bea
E_{\rm lc}=\mu\qquad
\alpha_0^{\dagger\, i}\, |0,p^+\rangle \,\,&\entspricht& 
\,\,\ft{1}{\sqrt{N^J}}\, \Tr\, (\phi_i\, Z^J) \label{eins}\\
E_{\rm lc}=\mu\qquad
\alpha_0^{\dagger\, \mu}\, |0,p^+\rangle \,\,&\entspricht& 
\,\,\ft{1}{\sqrt{N^J}}\, \Tr\, (D_\mu Z\, Z^{J-1})\label{zwei}\\
E_{\rm lc}=\mu\qquad
\theta_{0\, A}^{\dagger}\, |0,p^+\rangle \,\,\,&\entspricht& 
\,\,\ft{1}{\sqrt{N^J}}\, \Tr\, (\psi_A\, Z^{J})\label{dreis}
\eea
corresponding to the 8+8 bosonic and fermionic excitations
of the string. Note
that while $\Tr\, (\phi_i\, Z^J)$ is a conformal primary operator,
$\Tr\, (D_\mu Z\, Z^{J-1})$ is a descendant of the groundstate operator
$\Tr \, Z^J$ obtained by acting with $D_\mu$. Similarly
$\Tr\, (\psi_A\, Z\, Z^{J})$ is a superdescendant of the groundstate
operator\footnote{For a detailed discussion of BMN operators
and superconformal symmetry see \cite{Beisert:2002tn}.}. 
All three operators are 
again protected and have a total $\Delta-J=1$ exactly, matching the string
spectrum $E_{\rm lc}=\mu$. Higher zero mode excitation are modeled
by symmetrized insertions of $\phi_i$, $D_\mu Z$ and $\psi_A$ such as
\be
E_{\rm lc}=2\mu\qquad
\alpha_0^{\dagger\, i}\, \alpha_0^{\dagger\, j}\,
  |0,p^+\rangle \,\,\entspricht 
\,\,\ft{1}{\sqrt{J\, N^J}}\,\sum_{l=0}^J \Tr\, 
(\phi_i\, Z^l\, \phi_j\, Z^{J-l})
\qquad \Delta-J=2
\label{drei}
\ee
The generalization to higher modes should
be clear by now. 

Now obviously the true challenge lies in reproducing
the stringy mode spectrum through non-protected Super Yang-Mills
operators. At the first stringy mode level we are searching for 
a dual gauge theory object ${\cal O}^{ij}_n$ with
\be
E_{\rm lc}=2\mu\,\sqrt{1+\frac{n^2}{(\alpha'p^+\mu)^2}}\qquad
\alpha_{-n}^{i}\, \tilde\alpha_{-n}^{j}\,
  |0,p^+\rangle \,\,\entspricht 
\, {\cal O}^{ij}_n
\label{stringdict}
\ee
The operator ${\cal O}^{ij}_n$ should carry two impurities 
$\phi_i$ and $\phi_j$ in order to reproduce the $SO(4)$ index structure
of the dual string state. Moreover for $n\to 0$ it should reduce
to the protected operator of \eqn{drei}. The simplest ansatz is
\be
{\cal O}^{ij}_n=\ft{1}{\sqrt{J\, N^J}}\,\sum_{l=0}^J \Tr\, 
(\phi_i\, Z^l\, \phi_j\, Z^{J-l})\, f(n,l)
\ee
with a suitable function $f(n,l)$ obeying $f(0,l)=1$. It turns out that
the correct choice is 
\be
f(n,l)=e^{2\,\pi\,  i\, n\, l/J}
\label{corrch}
\ee
and we will
show in the next chapter why this is the case. Let us at this point just
state that a computation of the {\sl planar} scaling dimension 
with this choice of $f(n,l)$ up to one
loop order yields the result \cite{Berenstein:2002jq}
\be
\Delta_{{\cal O}^{ij}_n}= J+2+\frac{\gym^2\, N}{J^2}\, n^2 + 
{\cal O}(\gym^2)
\label{planardim}
\ee
Note the emergence of the promised effective coupling constant $\lambda'$ in
the BMN limit \eqn{BMNlimit}. In order to compare this result to
the string light-cone energy we need to convert the string parameters
$\alpha', p^+, \mu$ and $R$ to $\gym$ and $N$. From \eqn{lcscaling} 
we have
\be
2p^+=\frac{E+J}{\mu\, R^2}\sim \frac{2\, J}{\mu\, R^2}
\quad \Rightarrow \quad (\mu p^+)^2=\frac{J^2}{R^4}
\ee
which upon making use of the AdS/CFT relation $R^4=\gym^2\, 
\alpha^{\prime\, 2}\, N$ leads to
\be
\frac{1}{(\alpha'p^+\mu)^2}=\frac{\gym^2\, N}{J^2}=:\lambda'
\ee
Therefore the perturbative gauge theory expansion around $\lambda'=0$
corresponds to a $\mu\to\infty$ expansion on the string side, exactly
opposite to the flat Minkowski space regime $\mu=0$.

In this domain the square root
of the string state energy \eqn{stringdict} may be expanded out to
yield
\be
\frac{E_{\rm lc}}{\mu}= 2\, \sqrt{1+n^2\, \lambda'}= 2 +\lambda'\, n^2+
{\cal O}(\lambda^{\prime\, 2})
\ee
matching precisely with the Super Yang-Mills result \eqn{planardim}!
Recovering the complete structure of the free plane-wave string spectrum
therefore necessitates the summation of the complete planar perturbation
series on the gauge theory side. The consistency of the Super Yang-Mills
planar two loop result for the scaling dimension with 
the string theoretic square root was demonstrated in 
\cite{Gross:2002su} and a mechanism for the all loop result was
presented. Relying on certain assumptions a proof of the full square
root structure of the gauge theory scaling dimensions was obtained
thereafter in \cite{Santambrogio:2002sb}.

One therefore has strong evidence that the planar sector of BMN gauge
theory scaling dimensions indeed reproduces the free plane-wave
string spectrum, enabling one to write down a concrete dictionary linking
string states to gauge theory operators. 

\subsection{The non-planar sector of BMN gauge theory}

If the AdS/CFT duality is to hold in its strong version, i.e.
implying  the
exact equivalence of the full interacting $AdS_5\times S^5$ string 
theory to ${\cal N}=4$ Super Yang-Mills, one should be able to go 
beyond the free plane-wave string theory discussed above. 
How can one then recover the plane-wave string 
interactions on worldsheets of higher genera from the ${\cal N}=4$
gauge model in the BMN limit? The natural place to look for 
is the non-planar sector of the dual gauge theory, 
however, as was discussed in section one,  non-planar graphs are 
expected to be suppressed in the large $N$
limit. The surprising fact is that the BMN limit $N\sim J^2\to \infty$
represents a {\sl novel} double scaling limit of the gauge theory
under which graphs of all genera survive
\cite{Kristjansen:2002bb,Constable:2002hw}. Here the
suppression of non-planar graphs with $1/N^2$ is balanced by the
growing combinatorics of the diagrams involved with $J\to \infty$.
Next to the effective coupling constant $\lambda'$
a new {\sl effective} genus counting parameter $J^2/N$ arises
which remains finite and tunable in the BMN limit \eqn{BMNlimit}. Hence the
BMN gauge theory is controlled by two independent parameters
\be
\lambda':=\frac{\gym^2\, N}{J^2}\qquad \mbox{and}\qquad
g_2:=\frac{J^2}{N}
\ee
allowing for a double expansion.

The emergence of $g_2$ is most transparent in the
correlation function of two protected groundstate
operators $\Tr \, Z^J$, this being the simplest two-point function in
BMN gauge theory.
Here the exact coordinate and $\gym$ dependence
is trivial, as there are no loop-corrections,
and it remains to solve the combinatorial
problem of taking into account all possible contractions of
free field propagators. This may be efficiently summarized through a
correlator in a Gaussian complex matrix model,
\be
\langle \Tr\, Z^J(x)\, \Tr\, \bar Z^J(0)\rangle = 
\left( \frac{\gym^2}{8\pi^2|x|^2}\right )^J \,
\int dZ\, d\bar Z\,\Tr\, Z^J\, \Tr\bar Z^J\, e^{-\Tr ( Z\bar Z)} \, .
\label{MMintro}
\ee
Here the measure is an abbreviation of
\be
dZ\,d\bar Z= \prod_{a,b=1}^N\, \frac{d\mbox{Re}Z_{ab}\, 
d\mbox{Im}Z_{ab}}{\pi}
\qquad
\mbox{and}\qquad
\MMvev{{\cal O}}:= \int dZ\, d\bar Z\, 
{\cal O}\, e^{-\Tr (Z\bar Z)}
\ee
ensuring $\langle\, 1\, \rangle_{\rm M\, M}=1$. This matrix model
captures the correct gauge theory combinatorics by virtue of its propagator
\be
\MMvev{Z_{ab}\, \bar Z_{cd}} =\delta_{ad}\,\delta_{bc} \, ,
\ee
compare \eqn{prop2}.  As a matter of fact the correlator 
$\MMvev{\Tr Z^J\, \Tr \bar Z^J}$ can be computed exactly 
for finite $N$ and $J$ using matrix model techniques 
\cite{Kristjansen:2002bb}.
The result may be expanded as a series in $\frac{1}{N^2}$ 
and one extracts, for general $J$, the
corrections to the (trivial) planar result $J N^J$:
\begin{eqnarray}
\langle{\rm Tr}Z^J~ {\rm Tr}\bar Z^J \rangle_{\rm M\, M} & =&
J~N^J  \Bigg\{ 1 + \Bigg[ {J \choose 4} + {J \choose 3}\Bigg]~{1
  \over N^2} \cr 
& & + \Bigg[ 21 {J \choose 8}+49 {J \choose 7}
+36 {J \choose 6} + 8 {J\choose 5} 
\Bigg] {1 \over N^4} + \ldots 
\Bigg\} 
\label{expansion}
\end{eqnarray}

\begin{figure}[t]
\begin{center}
\begin{tabular}{ccc}
\epsfysize=3.5cm\epsfbox{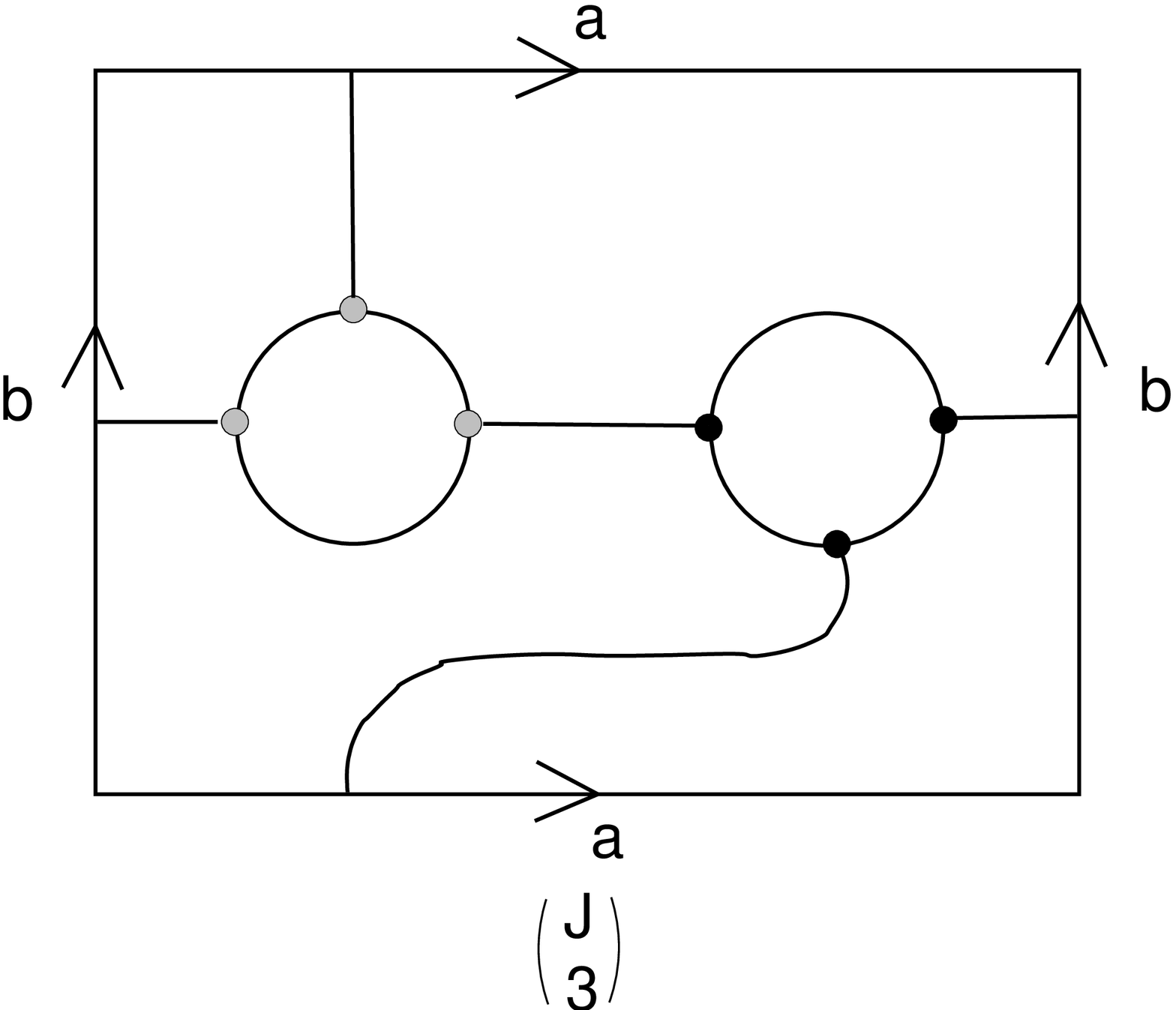}&\qquad&
\epsfysize=3.5cm\epsfbox{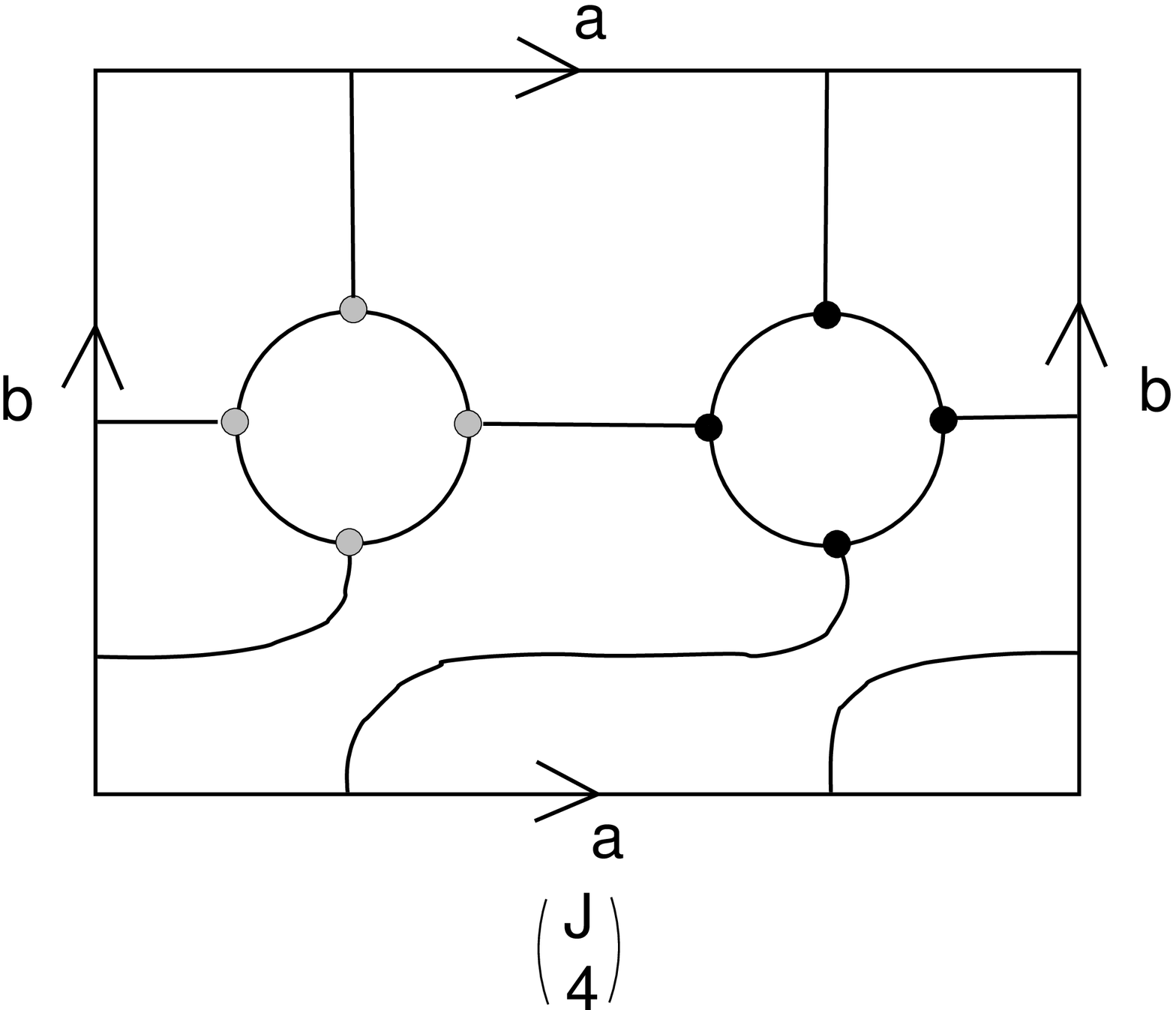}
\end{tabular}
\end{center}
\caption{The irreducible genus one graphs with their combinatorial
weight.}
\label{fig2}
\end{figure}

The structure of this result is easy to understand combinatorially.
We have to find the possible ways of connecting two necklaces with 
$J$ white ($Z$'s) and $J$ black ($\bar Z$'s)
beads respectively, according to the following
rules: (a) each connection has to link a black to a white bead,  
(b) in order to find the ${\cal O}(N^{J-2 h})$ contribution the 
connections have to be drawn without crossing on a genus $h$ 
surface such that no
handle of the surface can be collapsed without pinching a connection.
Let us call all connections that run (possibly after topological
deformation) parallel to another connection ``reducible''. Eliminate all
reducible connections. This will lead to a number of inequivalent,
irreducible graphs on the genus $h$ surface. There are two such irreducible
graphs for the toroidal contribution 
carrying combinatorial weights indicated in figure \ref{fig2} and 
appearing in
the terms of order $1/N^2$ in \eqn{expansion}. They arise from
distributing the $J$ beads of one necklace into three respectively four bins. 

Upon taking the BMN limit \eqn{BMNlimit} of the correlator 
\eqn{expansion} one is left with
\be
\frac{1}{J\, N^J}\, \langle{\rm Tr}Z^J~ {\rm Tr}\bar Z^J \rangle_{\rm M\, M}
\stackrel{J,N\to\infty} {\longrightarrow}
1 +\frac{1}{24}\, \frac{J^4}{N^2}+ \frac{21}{8!} \, \frac{J^8}{N^4}+
\ldots
= \frac{2\, N}{J^2}\, \sinh (\frac{1}{2}\, \frac{J^2}{N})
\ee
where we have inserted the all genus result obtained in
 \cite{Kristjansen:2002bb} in the last step. The central observation
here is that non-planar graphs are not fully suppressed
in this novel type of large $N$ limit and the new {\sl effective} 
genus counting parameter $g_2:=J^2/N$ arises. The class of non-planar
graphs surviving the BMN limit is distinctively smaller than the
class of all non-planar graphs, as is seen from the above toroidal
example. Only graphs with ``handles'' made of sufficiently many
beads contribute in the double scaling limit \eqn{BMNlimit}. 
This structure gives rises to a discretized closed string
interpretation of the gauge theory. Here a necklace of $J$ 
beads corresponds to a closed
string made out of $J$ bits and the $J\to\infty$ limit is a continuum
limit from the discrete string point of view. String splitting is
suppressed with $\frac{1}{N^2}$. The scaling law $J^2\sim N$ ensures
that only worldsheets with ``fat'' handles, i.e.~made out of a sufficiently
large number of string bits, survive the double scaling limit.
Smooth, macroscopic surfaces are formed in this limiting
process in great similarity to the
double scaling limits of the ``old matrix models'' relevant to
describing two-dimensional quantum gravity \cite{oldmm}.

Let us also stress that the chosen scaling of $J$ and $N$ indeed
represents a delicate balance between string splitting and 
string bit proliferations: Had one chosen a scaling law as
$J^p\sim N$ in the large $N$ limit, then for $p<2$ non-planar
graphs would actually dominate over planar ones, whereas
for $p>2$ only planar graphs would have survived. The BMN limit
choice $p=2$ is hence fine tuned to yield a full fleshed genus expansion
and justifies the name ``double scaling limit''.

\section{The computation of planar and non-planar corrections to
Super Yang-Mills scaling dimensions}

\begin{figure}[t]
\begin{center}
\begin{tabular}{ccc}
\epsfysize=2cm\epsfbox{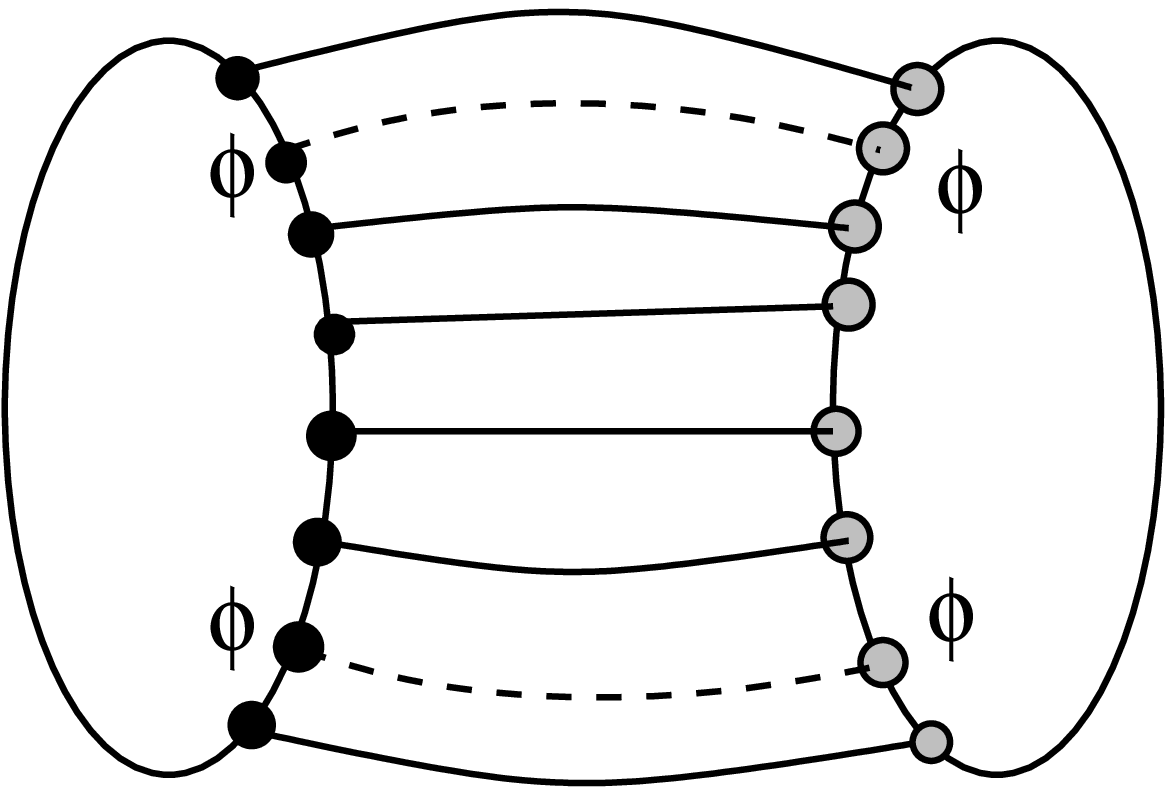} & \qquad & 
\epsfysize=2cm\epsfbox{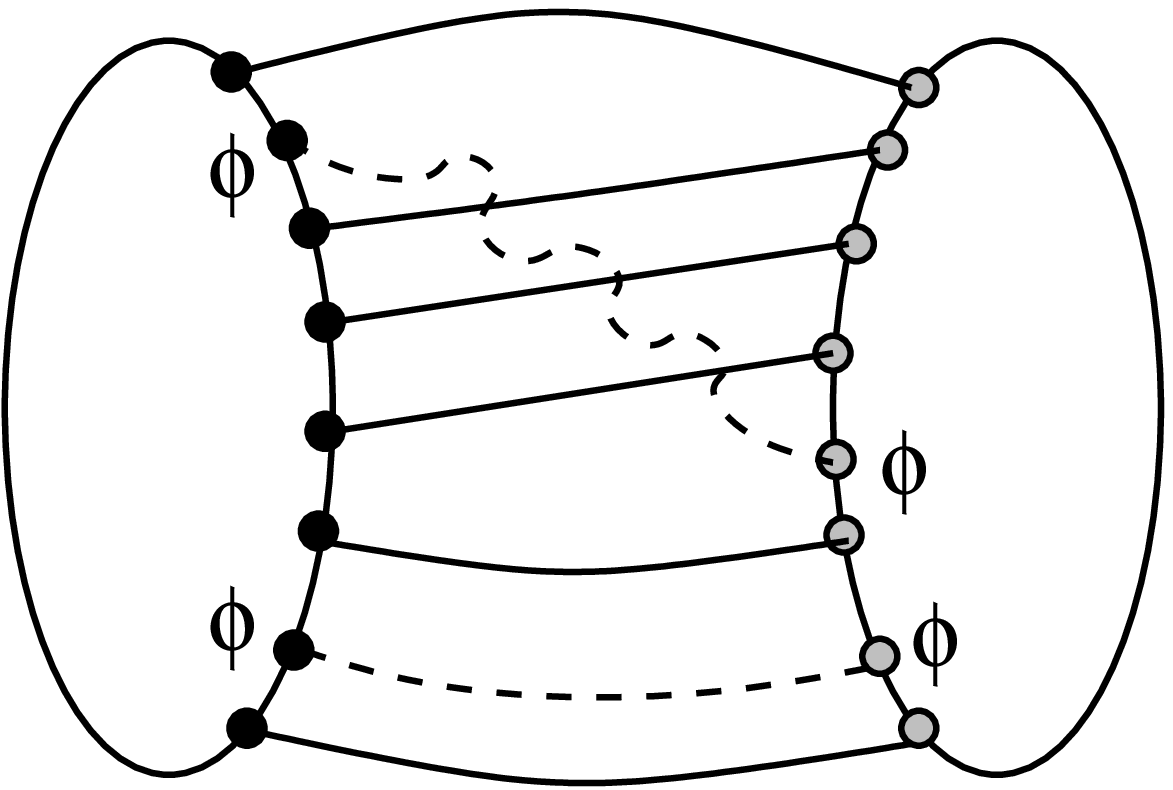}
\cr
Planar & & Non-planar \cr
$\cO^6_4 \to \cO^6_4$ && 
$\cO^6_4 \to \cO^6_1$ \cr
\end{tabular}
\end{center}
\caption{``Hopping'' is induced by non-planar graphs.}
\label{Fig3}
\end{figure}

After these general remarks and observations we now move on to the
explicit computation of scaling dimensions of BMN operators.
In these lectures we shall confine our attention to the 
scaling dimension of the two impurity single trace operator
\be
{\cal O}_p^J(x)= \Tr\, [\, \phi_1\, Z^p\, \phi_2\, Z^{J-p}\, ](x)
\ee
up to one quantum loop and genus one. For this we need to study the
two-point function
\be
\vev{\cO^J_p(x)\, \bar\cO^J_q(0)}= 
\vev{\cO^J_p(x)\, \bar\cO^J_q(0)}_{\rm classical}+
\vev{\cO^J_p(x)\, \bar\cO^J_q(0)}_{\rm 1-loop}\, .
\label{2pt1}
\ee
Here it will prove useful to again employ the matrix model techniques
encountered in our discussion of the protected operator $\Tr Z^J$. The
classical piece of \eqn{2pt1} is then
\be
\vev{\cO^J_p(x)\, \bar\cO^J_q(0)}_{\rm classical}=
\left (\frac{\gym^2}{8\pi^2|x|^2} \right )^{J+2}\,
\MMvev{\cO^J_p\, \bar\cO^J_q}
\ee
with 
\be
\MMvev{\cO^J_p\, \bar\cO^J_q}= \int dZ\, d\bar Z\, \Tr[Z^p\, \bar Z^{q}]
\, \Tr [ \, Z^{J-p}\, \bar Z^{J-q}]\, e^{-\Tr\, (Z\bar Z)}
\label{Mpq}
\ee
where we have contracted the two scalars $\phi_1$ and $\phi_2$ by making
use of the $U(N)$ fission and fusion rules
\be
\Tr[\undersym{\phi^+\, A\, \phi}^-\, B]=\Tr[ A]\,\Tr[ B]
\qquad
\Tr[\undersym{\phi^+\, A]\, \Tr [ \phi}^-\, B]=\Tr[ A\, B]
\label{fissfus}
\ee
with $\langle \undersym{\phi^+_{ab}\, \phi}{}^-_{cd} \rangle = \delta_{bc}\,
\delta_{ad}$,
an immediate consequence of \eqn{prop2}. The remaining 
correlator \eqn{Mpq} has been worked out explicitly in 
\cite{Kristjansen:2002bb,Constable:2002hw} up to genus one. 
One finds\footnote{This formula is valid only for
$(q>p\, , \,J-q>p)$ in the non-diagonal part.
The other regions of $p$ and $q$ are determined
by the two obvious symmetries $(p\leftrightarrow q)$ and 
$(p\rightarrow J-p, q\rightarrow J-q)$ of the correlator.}
\bea
\lefteqn{\MMvev{\cO^J_p\, \bar\cO^J_q}=
\MMvev{\Tr[Z^p\, \bar Z^{q}]\, \Tr [ \, Z^{J-p}\, \bar Z^{J-q}]}
=}\nn\\&\delta_{p,q}\, 
N^{J+2} +
N^J\, \Bigl [\delta_{p,q}\left[ {J-p+2\choose 4}+ {p+2\choose 4}\right]
+
\, \frac 1 6\, p\, (p+1)\,(3J+1-p-3q)& \nn\\&
+
(q-p)\, (p+1)\, (J-q+1)\, \Bigr ] \, +\, \cO(N^{J-2})&
\label{Jpq}
\eea
For higher genus results see \cite{Eynard:2002df}.
The important thing to note here is not the precise form of
this correlator but rather the fact that the classical
contribution to the two-point function
\eqn{2pt1} becomes non-diagonal in $p$ and $q$ at the toroidal level, 
i.e.~the non-diagonal $\cO(N^J)$ terms in
the above. The source of this non-diagonality are the non-planar
``hopping'' graphs depicted in figure \ref{Fig3}. 

Moving on to the one loop radiative corrections to the correlator
we need to include the scalar four point interaction, one loop self energy 
insertion and gluon exchange graph,
\be
\epsfysize=2.2cm\epsfbox{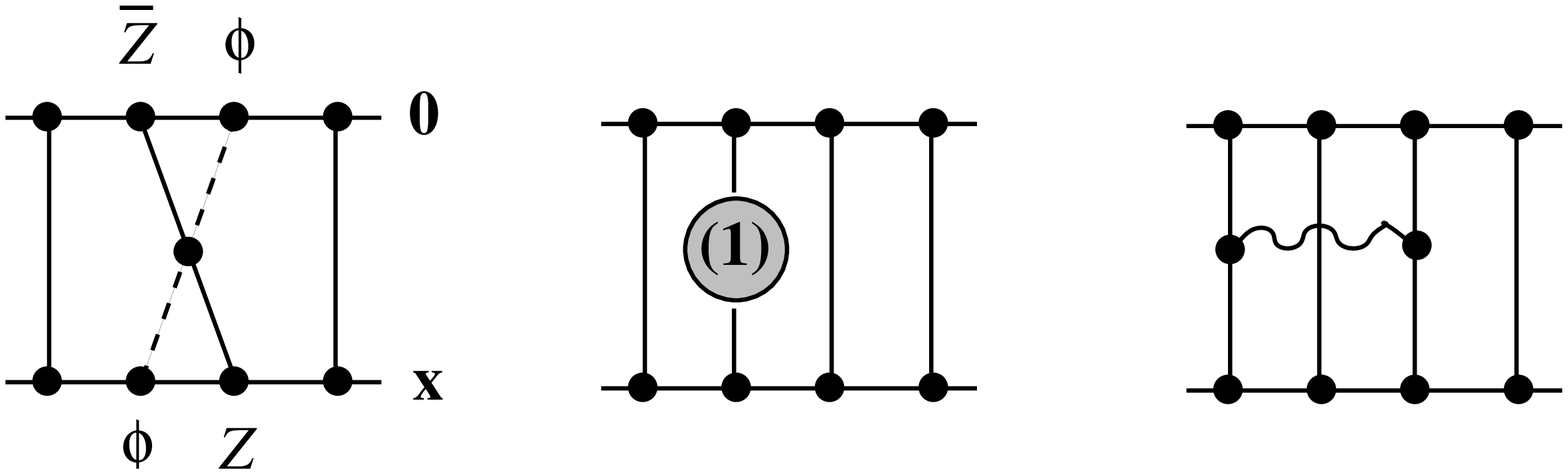}
\label{loops}
\ee
in the diagrams of figure \ref{Fig3}. Now the ``hopping'' already
occurs at the planar level, as seen in the above first diagram, which
exchanges the position of a $Z$ and $\phi_i$  field through 
the scalar self interaction term $\Tr [\phi_i,Z]\, [\phi_i,\bar Z]$
of the Lagrangian \eqn{N4SYM}. Therefore
\be
\vev{\cO^J_p(x)\, \bar\cO^J_q(0)}_{\rm 1-loop} \neq \delta_{p,q}
\label{1loopcorr}
\ee
and one needs to rediagonalize the set of BMN operators $\cO^J_p(x)$
in order to determine their scaling dimension from \eqn{CFT2PtFct}
already at the planar level.

How can one now incorporate the loop effects \eqn{loops} in an effective
matrix model vertex? As the matrix model is a zero dimensional
field theory we need to keep track of the positions ($0$ or $x$)
of the individual fields in the two-point function in form of two 
distinct matrix model fields. This was automatically
achieved via complexification for $Z$ (sitting at $x$) 
and $\bar Z$ (sitting at $0$), for the remaining four real scalar fields
$\phi_i$, however, we store this information in two different matrix
model fields denoted by $\Phi_i^\pm$
\be
\phi_i(0) \to \Phi_i^- \qquad
\phi_i(x) \to \Phi_i^+
\ee
with zero dimensional ``propagators''
\be
\MMvev{(\Phi_i^-)_{ab}\, (\Phi_j^+)_{cd}}= \delta_{ad}\, \delta_{bc}\,
\delta_{ij}\qquad
\MMvev{ \Phi_i^+\, \Phi_j^+}=0= 
\MMvev{ \Phi_i^-\, \Phi_j^-} \, .
\ee
In doing so we are able to  disentangle the space-dependence of the
correlation function \eqn{1loopcorr}
from the more challenging combinatorics. The combinatorics
of the scalar self interaction insertion is then given by the
effective matrix model vertex
\be
\ft 1 2\Tr[\phi_i,\phi_j]^2(x) \to
\Tr[\Phi_i^+,\Phi_j^-]\, [\Phi_i^+,\Phi_j^-] + 
\Tr [\Phi_i^+,\Phi_j^-]\, [\Phi_i^-,\Phi_j^+] + 
\Tr [\Phi_i^+,\Phi_j^+]\, [\Phi_i^-,\Phi_j^-] 
\ee 
as two fields need to be contracted with the operator at $0$
and the other two with the operator at $x$. Note that in the above
we have temporarily reverted to the index range $i=1,\ldots,6$. 
By making use of
the Jacobi identity this vertex may be reorganized into the more 
convenient form
\be
\ft 1 2\Tr[\phi_i,\phi_j]^2(x) \to V_D+ V_F +V_k
\ee
where
\bea
V_D &=& \phantom{-}\ft 12 \Tr[\Phi^+_i,\Phi_i^-]\, [\Phi^+_j,\Phi_j^-]
\qquad \mbox{symmetric piece}\nn\\
V_F &=& \,\,\,-\Tr[\Phi^+_i,\Phi_j^+]\, [\Phi^-_i,\Phi_j^-]
\qquad \mbox{anti-symmetric piece}\nn\\
V_K &=& -\ft 1 2 \Tr[\Phi^+_i,\Phi_j^-]\, [\Phi^+_i,\Phi_j^-]
\qquad \mbox{trace piece}
\eea
couple to the symmetric, anti-symmetric and trace pieces respectively
of an operator made entirely from the minus-valued matrix 
fields $\Phi_k^-$ upon contraction with the rules \eqn{fissfus}. 
In order to reproduce the field theoretic result this effective matrix 
model vertex needs to be augmented by a space dependent factor
arising  from the logarithmically divergent integral
\be
\raisebox{-0.45cm}{\epsfxsize=1.5cm\epsfbox{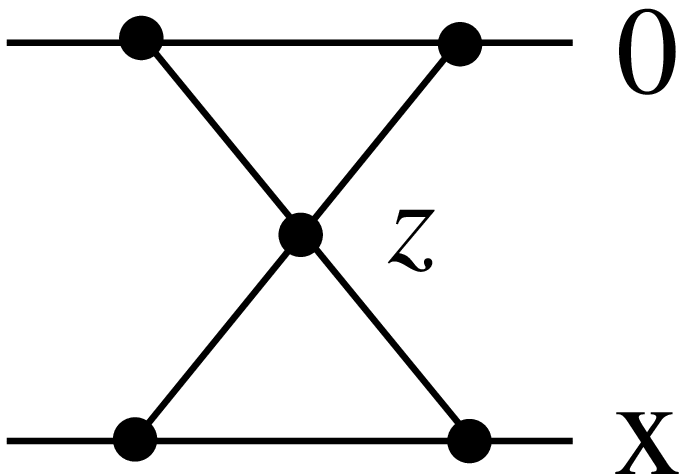}}
= -\frac{\gym^4\, x^4}{64\pi^4}\, \int \frac{d^4z}{(z-x)^4\, z^4}
\, = \, \frac{\gym^4\, L}{32 \pi^2}
\label{spacedepfact}
\ee
with
\be
L:= \log x^{-2} -(\ft 1 \epsilon + \gamma +\log \pi+2)
\ee
in dimensional regularization. This logarithmic dependence on $x$
is consistent with the one loop anomalous dimension $\Delta_1$
term arising in the $\lambda'\ll 1$ expansion of the conformal
field theory two-point function \eqn{CFT2PtFct}
\be
\frac{1}{x^{2({\Delta_0+\Delta_1})}}= \frac{1}{x^{2\,{\Delta_0}}}
\, \left ( 1 + {\Delta_1}\, \log(\Lambda\, x)^{-2} + \cO(\lambda'^2)
\right )
\ee
where we have introduced the scale factor $\Lambda$. 
The divergent piece of the integral
\eqn{spacedepfact} is canceled by an appropriate renormalization
of the operators $\cO^J_p(x)$. Hence, although the coupling constant
$\gym$ is not renormalized in the ${\cal N}=4$ model, composite
operators like $\cO^J_p(x)$ are.

In summary the effective matrix model vertex reflecting the 
scalar self interaction takes the form
\be
\Bigl (\,\raisebox{-0.3cm}{\epsfysize=0.8cm\epsfbox{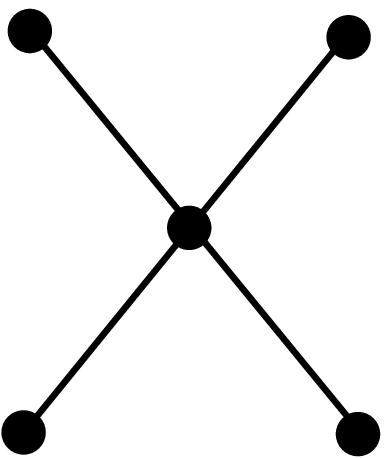}}
\, \Bigr )_{\rm {Matrix\, Model}}
= \frac{g^2_{\rm YM}\, L}{16\pi^2}\, \left (  :V_D: + :V_F: + :V_K: \right )
\ee
to be inserted in a matrix model correlator. Here the colons ``$:$''
denote normal ordering, disallowing self-contractions of fields within 
one vertex. The vertices for scalar self-energy
\begin{equation}
\Bigl (\,\raisebox{-0.4cm}{\epsfysize=0.9cm\epsfbox{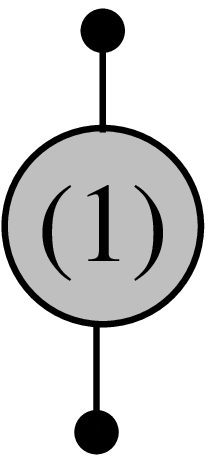}}
\, \Bigr )_{\rm {Matrix\, Model}}=
\frac{\gym^2 (L+1)}{8\pi^2}
 \Bigl (N\, :\Tr (\Phi^-_i\Phi^+_i):-:\Tr \Phi^-_i \,\Tr \Phi^+_i:\, 
\Bigl )
\end{equation}
and gluon-exchange
\begin{equation}
\Bigl (\,\raisebox{-0.4cm}{\epsfysize=0.9cm\epsfbox{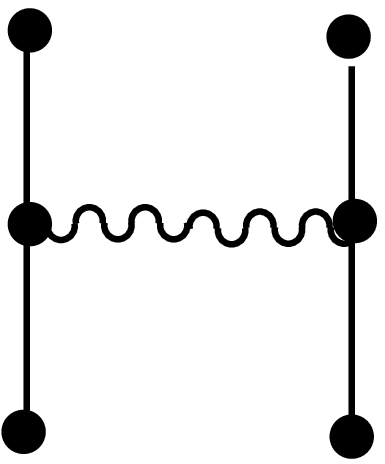}}
\, \Bigr )_{\rm {Matrix\, Model}}=
\frac{\gym^2(L+2)}{32\pi^2} \Bigl (\, 
:\Tr [\Phi^+_i,\Phi^-_i][\Phi^+_j,\Phi^-_j]:\, \Bigr ),
\end{equation}
are obtained in a similar fashion \cite{Beisert:2002bb}.   

Remarkably the term $V_D$ in the scalar interaction
cancels against the gluon-exchanges and
the scalar self-energies \cite{Constable:2002hw,Beisert:2002bb}.   
The proof goes as follows. 
The sum of these terms can be written 
without normal-orderings in the following way
\bea \label{D-terms}
\lefteqn{\left ( :V_D: + \, 
\raisebox{-0.35cm}{\epsfysize=0.9cm\epsfbox{se.eps}}
\, +\, 
\raisebox{-0.35cm}{\epsfysize=0.9cm\epsfbox{gluon.eps}} \, \right )_{\rm Matrix
\, Model}}\\
&=&\frac{\gym^2 (L+1)}{8\pi^2}
(\ft 12 \Tr[\Phi^+_i,\Phi^-_i][\Phi^+_j,\Phi^-_j]
-N\Tr \Phi^+_j \Phi^-_j+\Tr \Phi^+_j\,\Tr\Phi^-_j
).\nn
\eea
It is easy to see that the contraction of $\Phi^+_i$ in the above 
vertex with an arbitrary trace
of scalars $\Phi^-_{i_k}$ vanishes
\bea
\lefteqn{\Bigl (\Tr ([\Phi^+_i,\Phi^-_i][\Phi^+_j,\Phi^-_j])\,
\Tr (\Phi^-_{i_1} \Phi^-_{i_2}\cdots \Phi^-_{i_n})\Bigr )
_{\Phi^+_i\leftrightarrow \Phi^-_{i_k}}}
\nn\\
&=&\Tr (\, \Bigl [\, \Phi^-_{i_1},[\Phi^+_j,\Phi^-_j]\, \Bigr ]
\Phi^-_{i_2}\cdots \Phi^-_{i_n}\,)
+\Tr \Bigl (\Phi^-_{i_1} \Bigl [\Phi^-_{i_2}, [\Phi^+_j,\Phi^-_j]\Bigr ]
\Phi^-_{i_3}\cdots \Phi^-_{i_n}\Bigr )
+\ldots\nn\\
&&
+\Tr \Bigl (\Phi^-_{i_1}\cdots \Phi^-_{i_{n-1}}
\Bigl [\Phi^-_{i_n}, [\Phi^+_j,\Phi^-_j]\Bigr ]\Bigr ) 
\nn\\
&=& -\Tr \Bigl ([\Phi^+_j,\Phi^-_j] \Phi^-_{i_1}\Phi^-_{i_2}\cdots \Phi^-_{i_n}
\Bigr )
+\Tr \Bigl (\Phi^-_{i_1} \Phi^-_{i_2}\cdots 
\Phi^-_{i_n}[\Phi^+_j,\Phi^-_j]\Bigr )=0.
\eea
due to a telescoping sum and cyclicity of the trace. 
Furthermore, terms resulting from contracting the $\Phi_i^+$ with one of
the $\Phi^-_k$ inside the same vertex cancel against the remaining 
quadratic terms
in~\eqref{D-terms}. Thus the combination~\eqref{D-terms} does not give
any contribution to  two-point correlators of scalar fields.

In summary the total one-loop contribution to {\sl any} scalar 2-point
function in ${\cal N}=4$ Super Yang-Mills may be obtained through the
insertion of a simple effective matrix model vertex into a Gaussian
matrix model correlator
\be
\vev{\cO_1(x)\, \bar\cO_2(0)}_{\rm 1-loop} =
\frac{\gym^2\, L}{16\, \pi^2}\, \Bigl\langle
\cO_1^+\, (:V_F:+:V_K:)\,\bar\cO_2^-\Bigr \rangle_{\rm M\, M}\, .
\label{MMoneloop}
\ee
An immediate consequence of this result is the quoted 
non-renormalization of the BPS operators \eqn{CPO}: Due
to their symmetric/traceless structure they do not couple to
$V_F$ or $V_K$ and therefore the one-loop contribution to
{\sl any} two-point function involving a chiral primary
operator $\cO_{\rm CPO}$ \eqn{CPO} vanishes.
Recently this construction has been generalized to the case
of an arbitrary (scalar, vector or spinor) two-point function
in ${\cal N}=4$ Super Yang-Mills in \cite{Beisert:2003jj}.

Let us now apply this insight to the computation of our two-impurity BMN
operator $\cO^J_p:= \Tr(\phi_1Z^p\phi_2 Z^{J-p})$ where we were
interested in
\be
\vev{\cO^J_p(x)\, \bar\cO^J_q(0)}= 
\left (\frac{\gym^2}{8\pi^2|x|^2} \right )^{J+2}\,
\Bigl (\, S_{pq} + T_{pq}\, \log |x\, \Lambda|^{-2}\, \Bigr )
\label{2PtFct}
\ee
with $S_{pq}=\MMvev{\cO^J_p\bar \cO^J_q}$ given in \eqn{Jpq} and 
\be
T_{pq}=-\ft{\gym^2}{8\pi^2}\,\Bigl\langle
\Tr(\phi_1^+ Z^p \phi_2^+ Z^{J-p})\,
:\Tr[\bar Z,\phi_i^-]\, [Z,\phi_i^+]:\,
\Tr(\phi_1^-\bar Z^p \phi_2^- \bar Z^{J-p})\,\Bigr\rangle_{\rm M\, M} \nn
\label{Tpq}
\ee
having inserted the relevant terms of $V_F$ in the 
$(Z,\bar Z,\phi_{i=1,2,3,4})$ basis. Note that the trace piece $V_K$ of the
effective vertex does not contribute here. Evaluating this correlator
is a straightforward yet tedious problem and we shall see soon that
this is actually not necessary for the determination of $\Delta$.

\begin{figure}[t]
\begin{center}
\begin{tabular}{c}
\epsfysize=2cm\epsfbox{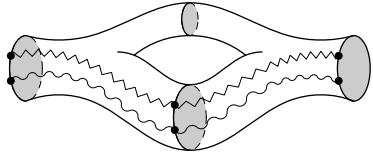}
\end{tabular}
\end{center}
\caption{Cutting the torus of a correlator between two single trace operators
yields double trace operators. The wiggly lines represent the impurity
insertions.}
\label{Fig4}
\end{figure}

Before proceeding, there is an additional complication we have to face, known
as operator mixing \cite{Bianchi:2002rw}. In principle one could diagonalize 
the two-point functions of the $\cO^J_p(x)$ 
of \eqn{2PtFct}, this, however,  is {\sl not}
correct. The reason is easy to understand pictorially: Considering the
torus correction to a two-point function, we see from figure \ref{Fig4}
that double-trace operators appear in intermediate channels. And indeed
the overlap between such double-trace operators and the single-trace
BMN operators is of $\cO(g_2)$. It therefore affects the $\cO(g_2{}^2)$
anomalous dimension upon diagonalization
\cite{Beisert:2002bb,Constable:2002vq}. 
We conclude that we have to
consider the enlarged set of multi-trace BMN-operators of the form
\be
\cO^{J_0,J_1,\ldots, J_k}_p = 
\Tr\, [\, \phi_1\, Z^p\, \phi_2\, Z^{J_0-p}\, ]\,
\Tr\, Z^{J_1}\ldots\Tr\, Z^{J_k}
\label{multi-trace-ops}
\ee
with $J_0+J_1+\ldots + J_k=J$ in our computation of two-point functions.
The reader might wonder why one does not
need to include a multi-trace operator with a single impurity insertion
in the first and the second trace of the form
\be\tilde\cO^{J_0,J_1,\ldots, J_k}=
\Tr\, [\, \phi_1\, Z^{J_0}\, ]\,
\Tr\, [\, \phi_2 Z^{J_1}\, ] \, \Tr Z^{J_2}\ldots\Tr\, Z^{J_k}
\; .
\ee
We shall see in the following that there is not mixing with
these types of states\footnote{The reason for this is that 
$\tilde\cO^{J_0,J_1,\ldots, J_k}$ is a protected operator, which does
not mix with the generically unprotected operators 
$\cO^{J_0,J_1,\ldots, J_k}_p$ of 
\eqn{multi-trace-ops}.}.

\subsection{BMN-gauge theory as a quantum mechanical system}

We will now review the paper \cite{Beisert:2002ff} in which an
efficient and simple method for the computation of anomalous
contributions to the scaling dimension $\Delta$ was introduced.
For the general mixing situation at hand we need to study the two-point
functions of multi-trace operators up to (say) one-quantum loop
\be
\vev{\cO_\alpha(x)\, \bar\cO_\beta(0)}= 
\left (\frac{\gym^2}{8\pi^2|x|^2} \right )^{J+2}\,
\Bigl (\, S_{\alpha\beta} + T_{\alpha\beta}\, \log |x\, \Lambda|^{-2}\, \Bigr )
\label{Gen2PtFct}
\ee
with $\alpha,\beta$ being multi-indices running over the set of 
multi-trace operators introduced in
\eqn{multi-trace-ops}. $S_{\alpha\beta}$ denotes the tree-level and
$T_{\alpha\beta}$ the one-loop matrix model correlators
\be
S_{\alpha\beta}=\MMvev{\cO_\alpha\,\bar \cO_\beta}\qquad
\mbox{and}\qquad
T_{\alpha\beta}=\MMvev{\cO_\alpha\,H\,\bar \cO_\beta}
\ee
with $H:=-\frac{\gym^2}{8\pi^2}\, :\Tr[Z,\Phi_i^+][\bar Z,\Phi_i^-]:$.
We are seeking a basis transformation to a new set of operators $\cO'_A$
\be
\cO_\alpha =V_{\alpha A}\, \cO'_A
\label{VaA}
\ee
which possess definite scaling dimensions $\Delta_A$ read off from the
{\sl diagonal} two-point functions
\be
\vev{\cO'_A(x)\,\bar\cO'_B(0)}=\frac{\delta_{AB}\, c_A}{|x|^{2(J+2+\Delta_A)}}
\, ,
\ee
where $c_A$ is a normalization constant. Put differently the operators 
$\cO_A'$ are eigenstates of the anomalous piece of the 
dilatation operator ${\cal D}$ with eigenvalues $\Delta_A$, 
i.e.~${\cal D}\circ \cO'_A= \Delta_A\, \cO'_A$.
Re-expressed in the original basis we have
\be
{\cal D}\circ \cO_\alpha = (V_{\alpha A}\, \Delta_A\, V^{-1}_{A\beta})\,
\cO_\beta
\label{Ddef}
\ee
giving us the matrix elements ${\cal D}_{\alpha\beta}$
of the anomalous piece of the
dilatation operator in the original,
non-diagonal basis. Hence
\bea
\vev{\cO_\alpha(x)\, \bar\cO_\beta(0)}&=&
V_{\alpha A}\, V^\ast_{\beta B}\, \vev{\cO_A'(x)\, \bar\cO_B'(0)}=
V_{\alpha A}\, V^\ast_{\beta B}
\, \frac{\delta_{AB}\, c_A}{|x|^{2(J+2+\Delta_A)}}\nn\\
&=&
V_{\alpha A}\, V^\ast_{\beta B}
\, \frac{\delta_{AB}\, c_A}{|x|^{2(J+2)}}\Bigl ( 1 +\Delta_A\, 
\log |x\Lambda|^{-2}\Bigr )\nn\\
&=& \frac{1}{|x|^{2(J+2)}}\, \Bigl [ 
\underbrace{(V\, C\, V^\dagger)}_{S_{\alpha\beta}}{}_{\alpha\beta}
+ \underbrace{(V\, C\, 
\Delta\, V^\dagger)}_{T_{\alpha\beta}}{}_{\alpha\beta}
\, \log |x\Lambda|^{-2}\Bigr ]
\eea
where we introduced the diagonal matrices $C_{AB}=\delta_{AB}\, c_A$
and $\Delta_{AB}=\delta_{AB}\, \Delta_A$.
 We conclude that the dilatation matrix
${\cal D}_{\alpha\beta}$ of \eqn{Ddef} may be expressed as \cite{Janik:2002bd}
\be
T_{\alpha\gamma}\, (S^{-1})_{\gamma\beta}= ( V\, C\,\Delta\, V^\dagger 
\, V^{\dagger\, -1}\, C^{-1}\, V^{-1})_{\alpha\beta}
= (V\,\Delta\, V^{-1})_{\alpha\beta}={\cal D}_{\alpha\beta}\, ,
\label{TDid}
\ee
the diagonal matrices $\Delta$ and $C$ commute. Now consider the
action of the effective vertex (or Hamiltonian) 
$H=-\frac{\gym^2}{8\pi^2}\, :\Tr[Z,\Phi_i^+][\bar Z,\Phi_i^-]:$
on the left-hand-side operator (or state) $\cO_\alpha(Z,\Phi^+_i)$
by Wick contraction
\be
H\circ \cO_\alpha = H_{\alpha\beta}\, \cO_\beta
\qquad \mbox{with} \qquad H_{\alpha\beta}: \mbox{$c$-numbers}\, .
\ee
This relation is to be understood as follows. Contract the two
minus-indexed fields $\bar Z$ and $\Phi^-_i$ of $H$ with
the plus-indexed fields of the operator $\cO_\alpha(Z,\Phi^+_i)$
according to the rules \eqn{fissfus}. The result
will be a linear combination of operators of the type 
$\cO_\alpha(Z,\Phi^+_i)$ again with coefficients $H_{\alpha\beta}$.
In the correlation function one then simply has
\be
T_{\alpha\beta}= \MMvev{H\circ \cO_\alpha\, \bar\cO_\beta}=
H_{\alpha\gamma}\, \MMvev{\cO_\gamma\, \bar\cO_\beta}=
H_{\alpha\gamma}\, S_{\gamma\beta}
\label{THS}
\ee
and therefore the one-loop anomalous piece of the dilatation operator
in the non-diagonal basis
is simply given by the matrix elements $H_{\alpha\beta}$ upon using
\eqn{TDid}:
\be
D_{\alpha\beta}= (J+2)\, \delta_{\alpha\beta} + H_{\alpha\beta}
\ee
Hence, the knowledge of tree-level mixing matrix
$S_{\alpha\beta}$, whose involved structure as an infinite expansion
in $g_2$ we already encountered in \eqn{Jpq}, is {\sl completely 
unessential} for the determination of the anomalous dimensions 
$\Delta_A$! These simply correspond to the eigenvalues of the matrix
$H_{\alpha\beta}$. Let us note that while $T_{\alpha\beta}$ and
$S_{\alpha\beta}$ are Hermitian by definition, $H_{\alpha\beta}$
is {\sl not}, but rather
\be
H^\dagger = S^{-1}\, H\, S
\label{nonherm}
\ee
easily deduced from \eqn{THS}.

As we will discuss in the following section the dilatation 
operator with matrix elements 
$(J+2)\delta_{\alpha\beta}+H_{\alpha\beta}$ finds its natural
dual in the string field theory Hamiltonian $\widehat H_{\rm string}$
of the full interacting plane-wave string theory. One is lead to propose
the {\sl operator} correspondence \cite{Gross:2002mh}
\be
\frac{\widehat H_{\rm string}}{\mu} \entspricht 2\,\delta_{\alpha\beta}+
H_{\alpha\beta}
\ee
where the left-hand side acts in the interacting plane-wave
string theory and the right hand side acts in BMN gauge theory.

Let us now compute $H_{\alpha\beta}$. Acting on $H$ with the
single-trace operator $\cO^J_p$ yields again single-trace
operator terms of the same type, but also double-trace operators:
\bea
H\circ \cO^J_p&=&\ft{\gym^2}{4\pi^2}\,\Bigl [ N(2\, \cO^J_p-\cO^J_{p-1}-\cO^J_{p+1})\nn\\&& +
\sum_{l=1}^{p-1}\, \{ \cO^{J-l,l}_{p-l} - \cO^{J-l,l}_{p-l-1}\, \}
+ \sum_{l=1}^{J-p-1}\, \{ \cO^{J-l,l}_{p} - \cO^{J-l,l}_{p+1}\, \}
\, \Bigr ]
\label{H1}
\eea
This shows that the action of $H$ on single-trace operators does not close.
Acting with a double-trace operator leads to double, triple and
single trace operators
\bea
H\circ \cO^{J_0,J_1}_p&=&\ft{\gym^2}{4\pi^2}\,
\Bigl [ N(2\, \cO^{J_0,J_1}_p-\cO^{J_0,J_1}_{p-1}-\cO^{J_0,J_1}_{p+1})\nn\\&& +
\sum_{l=1}^{p-1}\, \{ \cO^{J_0-l,J_1,l}_{p-l} 
- \cO^{J_0-l,J_1,l}_{p-l-1}\, \}
+ \sum_{l=1}^{J_0-p-1}\, \{ \cO^{J_0-l,J_1,l}_{p} - 
\cO^{J_0-l,J_1,l}_{p+1}\, \} \nn\\&&
+ J_1\,( \cO^J_{J_1+p}-\cO^J_{J_1+p-1}+\cO^J_p-\cO^J_{p+1}\, )
\, \Bigr ]
\label{H2}
\eea
Here we have neglected boundary terms which are irrelevant in the
BMN limit. Hence the Hamiltonian
decomposes as $H=H_0+H_++H_-$ where $H_0$ maps $n$-trace operators
into $n$-trace operators and $H_+$ raises and $H_-$ lowers the
number of traces by one. Clearly the $H_\pm$ pieces 
correspond to string splitting and joining
interactions in a dual string model.

One can directly rephrase \eqn{H1} and \eqn{H2} in the BMN limit, amounting
to a continuum limit of the discretized string picture.
For this upon taking $J\to \infty$ one introduces the continuum  variables
\be
x:=\frac{p}{J}\quad\mbox{with $x\in[0,r_0]$;} \quad
r_i:=\frac{J_i}{J} \quad\mbox{with $r_i\in[0,1]$;} \quad
\mbox{and}\quad \sum_{l=0}^kr_l=1 \, .
\ee
Then the discrete multi-trace operators are replaced by continuum
states
\bea
\cO^J_p &\to & |x\rangle \nn\\
\cO^{J_0,J_1}_p &\to & |x;r_1\rangle \nn\\
\cO^{J_0,J_1,\ldots,J_k}_p &\to & |x;r_1,\ldots,r_k\rangle 
\eea
where $x$ may be interpreted as the coordinate of a particle moving
on a circle of circumference $r_0$, as one identifies
$|x;r_1,\ldots, r_k\rangle \sim |x+r_0;r_1,\ldots, r_k\rangle $. 
The discrete action of $H\circ \cO^J_p$
may now be rephrased as the action of a quantum mechanical Hamiltonian
on $|x;r_1,\ldots,r_k\rangle$:
\bea
\hat H\, |x\rangle &=& \frac{\lambda'}{4\pi^2}\, \Bigl [ -\partial_x^2|x\rangle
+g_2\int_0^xdr\, \partial_x|x-r;r\rangle -
g_2\int_0^{1-x}dr\, \partial_x|x;r\rangle \,
\Bigr ]
\label{con1}
\eea
from \eqn{H1} with the genus counting parameter $g_2=J^2/N$ appearing
and
\bea
\hat H\, |x;r_1\rangle &=& \frac{\lambda'}{4\pi^2}
\, \Bigl [ -\partial_x^2|x;r_1\rangle
+g_2\, r_1(\partial_x|x+r_1\rangle - \partial_x|x\rangle) \nn\\&&
+ g_2\int_0^xdr_2\, \partial_x|x-r_2;r_1,r_2\rangle -
g_2\int_0^{r_0-x}dr_2\, \partial_x|x;r_1,r_2\rangle
\,\Bigr ]
\label{con2}
\eea
from \eqn{H2}. The eigenvalues of $\hat H$ now correspond to the
anomalous dimensions of the BMN operators. Let us stress the
remarkable fact that the interaction terms of $\hat H$ terminate
at order $g_2$. If we were to succeed in diagonalizing $\hat H$
exactly, we would have found an all genus prediction for interacting
plane-wave string! Unfortunately one does not (yet) know how
to do this and we have to revert to perturbation theory in $g_2$. 

The planar ($g_2=0$) sector is easily diagonalized. We have
$\hat H_0=-\frac{\lambda'}{4\pi^2}\partial_x^2$ whose single-trace 
eigenstate reads (with $n$ integer due to periodicity)
\be
|n\rangle =\int_0^1dx\, e^{2\pi i\, n\, x}\, |x\rangle
\qquad
\mbox{with}
\qquad 
\hat H_0\, |n\rangle = \lambda'\, n^2\, |n\rangle\, .
\label{H01}
\ee 
This result yields the promised one-loop planar anomalous
dimension of the two impurity BMN operator $\Delta_1=\lambda'\, n^2$
as advertised in  \eqn{planardim}. It also shows that the
correct discrete BMN operator indeed is
\be
\alpha_{-n}^{i}\, \tilde\alpha_{-n}^{j}\,
  |0,p^+\rangle \quad\entspricht\quad
\cO_n=\sum_{l=0}^J e^{2\pi i\, n\, l /J}\, \cO^J_p
\ee
as claimed in \eqn{corrch}. The generalization of \eqn{H01} to
multi-trace eigenstates is obvious
$$
|n;r_1,\ldots, r_k\rangle = \ft{1}{\sqrt{r_0}}\, \int_0^{r_0}dx\, 
e^{2\pi i \, n\, x/r_0}\, |x; r_1,\ldots, r_k\rangle
$$
with 
\be
\hat H|n;r_1,\ldots,r_k\rangle = \lambda'\, \frac{n^2}{r_0^2}
\, |n;r_1,\ldots,r_k\rangle \, .
\label{H0k}
\ee 
Note that whereas the spectrum of single-trace states is
discrete, the spectrum of multi-trace states is continuous
due to $r_0\in [0,1]$.

Higher genus corrections to the anomalous scaling dimension
$\Delta$ may now be computed from simple quantum mechanical
perturbation theory. The genus one correction to \eqn{H01} reads
\be
E^{(1)}_{|n\rangle}= \langle n| \hat H_{\rm int}\, 
\frac{1}{E^{(0)}_{|n\rangle}-\hat H_0}\, \hat H_{\rm int}
|n\rangle
\qquad
\mbox{where}\qquad E^{(0)}_{|n\rangle}=\lambda'\, n^2
\label{SYMg1}
\ee
and where $\hat H_{\rm int}$ denote the $g_2$ dependent terms
in \eqn{con1} and \eqn{con2}. In order to perform this computation 
one needs the matrix element of $H_{\rm int}$ in the momentum
basis, 
\begin{subequations}\label{matel}
\begin{align}
\langle m,r| \hat H_{\rm int}|n\rangle &= - \frac{1}{(1-r)^{3/2}}
\, \frac{4m}{n-\ft{m}{1-r}}\, \sin^2 (\pi n r)\\
\langle n| \hat H_{\rm int}|m,r\rangle &= \frac{r}{(1-r)^{1/2}}
\, \frac{4n}{n-\ft{m}{1-r}}\, \sin^2 (\pi n r)\, . \label{nHmr}
\end{align}
\end{subequations}
Note the non-hermiticity of $\hat H_{\rm int}$ as discussed 
in \eqn{nonherm}. It is now a one line computation to compute
the genus one energy shift to be 
\be
E^{(1)}_{|n\rangle}
=\int_0^1 dr  \sum_{m=-\infty}^\infty 
\langle n|\hat H_{\rm int}|m;r\rangle \, 
\frac{1}{4\pi^2(n^2-\frac{m^2}{(1-r)^2})}\, \langle m;r|
\hat H_{\rm int}|n\rangle
= \frac{1}{12}+\frac{35}{32\,\pi^2n^2}\, .
\label{oneline}
\ee
Summarizing, the gauge theory prediction for the higher
genus corrections to the plane-wave string spectrum read
\bea
\lefteqn{
\frac{E_{|n\rangle}}{\mu} = 2+\lambda' \Bigl [ n^2+ g_2{}^2
\frac{1}{4\pi^2}\, (\frac{1}{12}+\frac{35}{32\pi^2n^2})
+}\nn\\&
 g_2{}^4\frac{1}{4\pi^2}\, ( -\frac{11}{46080}\,\,\frac{1}{\pi^2n^2 }
+\left(\frac{521}{12288}-\frac{\zeta(3)}{128}\right)\frac{1}{\pi^4 n^4}
+
\left(-\frac{5715}{16384} - \frac{45\,\zeta(3)}{512} +
  \frac{15\,\zeta(5)}{128}
\right)\frac{1}{\pi^6n^6})\, 
\Bigr ]& \label{gtg2}
\eea
where we have also spelled out the more involved result of the
genus two computation \cite{Beisert:2002ff}.
There is an important subtlety concerning the use of {\sl non}-degenerate
perturbation theory in the diagonalization of $H$. This is a
priori not justified as single-trace states $|n\rangle$ of \eqn{H01}
are degenerate in energy with multi-trace states 
$|m;s_1,\ldots, s_k\rangle$ of \eqn{H0k} for $n\cdot s_0=\pm m$
\footnote{Actually
the $n=1$ state is non-degenerate as $s_0=1$ would be required,
from which $s_i=0$ follows, turning the multi-trace state into a 
single-trace one. But $n=2$ is degenerate with $m=\pm 1$, $s_0=1/2$,
$n=3$ is degenerate with $m=\pm 2$, $s_0=2/3$ and $m=\pm 1$, $s_0=1/3$
and so forth.}. It turns out that this degeneracy does not
lead to problems in the genus one computation \eqn{oneline}, as
the overlaps of degenerate single with double-trace states
of \eqn{matel} vanish\footnote{I.e.~for degeneracy one
has $n\cdot (1-r)=m$ leading to poles in \eqn{matel}.
This however implies $n\cdot r$ being integer letting
the numerator vanish to yield a vanishing overlap.}. However, degeneracy
of single with triple-trace operators leads to a breakdown of
non-degenerate perturbation theory from genus two on.
This may be interpreted in the dual string theory
as an instability of excited single
string states to decay into the continuum
of degenerate triple-string states. See \cite{Freedman:2003bh} for
a discussion of this, a confirmation of the genus two result
\eqn{gtg2} and the computation of the corresponding
decay width. 

In the above we have only considered the insertion of two
scalar impurities of distinct type. The case of two
general scalar impurities \cite{Beisert:2002bb} (at planar and
toroidal level), mixed vector-scalar impurities \cite{Gursoy:2002yy} and 
two vector impurities \cite{Klose:2003tw} has been analyzed as
well, leading to identical planar anomalous dimension
$\Delta_0=\lambda'\cdot n^2$, in agreement with the 
free string spectrum. The necessity of this observed degeneracy
of general two impurity states (to all orders in $g_2$)
was proved in \cite{Beisert:2002tn} 
by employing the underlying superconformal symmetry.

Recently the structure of the dilatation operator at the
two loop level, i.e.~at order $\lambda'^2$ was
established \cite{Beisert:2003tq}. The result at two-loops
led the authors to conjecture that the Hamiltonian of
BMN gauge theory is given to all loop orders by 
\be
H_{\rm full} = 2\sqrt{1+\lambda'\, H}
\label{Hfull}
\ee
with $H=H_0+g_2\, H_++g_2\, H_-$ 
being the one-loop Hamiltonian discussed above. This
result is intriguingly simple and 
consistent with the planar, all-loop result
of Santambrogio and Zanon \cite{Santambrogio:2002sb} and agrees
with the free string spectrum. 
It is firm, however,
only at order $\lambda'^2$. Let us also note that the
diagonalization of $H_{\rm full}$ at higher orders in $\lambda'$
is thus reduced to the one loop diagonalization of $H$, which we
have performed perturbatively in \eqn{gtg2}. Knowing the
spectrum of the one-loop Hamiltonian
$H$ would yield the exact interacting string
spectrum to all orders in $g_2$ and $\lambda'$.

\section{Interacting plane-wave superstrings}

After having established the gauge theory predictions for higher
genus corrections to the plane-wave string spectrum let us 
now  study how these corrections
can be obtained in a string theory computation. For this one needs
to study string interactions arising from higher genus
worldsheets. It turns out that the standard vertex operator
methods for the computation of string scattering amplitudes
are not easily generalized to the plane-wave background. Instead the 
methods of light-cone string field theory, developed in 1973-75
\cite{Mandelstam}, have been successfully reformulated 
for the plane-wave superstring and used to compute the
genus one corrections to the spectrum. This subject 
is highly technical in nature and we
do not have the space in these lecture notes to develop it in
detail -- we have to refer the reader to the original
literature as we proceed, also see \cite{Pankiewicz:2003pg} for 
a recent review.  
Instead our emphasis in this section will be to display 
the key points and structural issues of this technically 
involved subject.

To begin with we shall rewrite the Hamiltonian \eqn{HLK1} of the
free, light-cone superstring in the plane-wave background 
in the following unified
notation \cite{Spradlin:2002ar,Pankiewicz:2002tg}
\be
H_2=p^-=\ft{1}{\alpha'\, p^+}\sum_{n\in \Z}\, \omega_n\, 
(a^{\dagger I}_n\,  a^I_n + b_n^{\dagger}\, b_n )
\qquad \mbox{where}\qquad
\omega_n:=\sqrt{n^2+(\alpha'p^+\mu)^2}
\label{HLK2}
\ee
with $[a_n^I, a_m^{\dagger J}]=\delta^{IJ}\, \delta_{n,m}$
and $\{b_n^a, b_m^{\dagger b}\}=
\delta^{ab}\, \delta_{n,m}$ where $a,b=1,\ldots, 8$. 
The $a^I_n$ oscillators
are related to the $\alpha^I_n$ and $\tilde\alpha^I_n$  oscillators
of section three via (suppressing the space index $I$)
\be
a_n^\dagger= \begin{cases}\alpha_{-n} & n>0 \cr
                    \tilde\alpha_{-|n|} & n<0 \cr
                    \alpha^\dagger_{0} & n=0 
\end{cases}
\qquad\qquad
a_n= \begin{cases}\alpha_{n} & n>0 \cr
                    \tilde\alpha_{|n|} & n<0 \cr
                    \alpha_{0} & n=0 
\end{cases}
\ee
In a similar fashion the fermionic modes $\theta^{(1,2)}_n$
of \eqn{HLK1} combine into the complex fermionic oscillators
$b^\dagger_m$ and $b_m$. The single string Hilbert space is then
built on the vacuum-state $|0\rangle$ subject to
\be
a^I_n|0\rangle=0 \qquad b_n|0\rangle=0 \qquad n\in \Z
\label{sftvacdef}
\ee
and physical states have to satisfy the level-matching
condition
\be
\sum_{n\in \Z} n (a^{I \dagger}_n\, a^I_n +b^\dagger_n\, b_n )\,
|\mbox{phys}\rangle =0 \, .
\ee

A central role in the construction of the interacting string field
theory is
the structure of the plane-wave superalgebra, which may be obtained
by a contraction of the $AdS_5\times S^5$ symmetry algebra
as a consequence of the Penrose limit.
The isometries of the plane-wave metric \eqn{pwmetric} are generated
by the Hamiltonian $H=P^-$ and 
momentum operators $P^I$, $P^+$ as well as the
angular momentum operators $J^{+I}$, $J^{ij}$ and $J^{i'j'}$, where
we denote $I=(i,i')$ with $i=1,2,3,4$ and $i'=5,6,7,8$. 
Let us stress that there are no isometry generators
$J^{-+}$ and $J^{-I}$ present in the algebra, a manifestation of
the broken Lorentz symmetry in the plane-wave background.
Additionally
the 32 supersymmetries are generated by the supercharges $Q^+$
and $Q^-$. Explicitly these read \cite{Metsaev:2001bj} (at $\tau=0$)
\be
Q^+ = \int d\sigma\, \sqrt{2}\, \lambda \qquad
Q^- = \int d\sigma\, [2\pi\alpha'\slashed{p}\, \lambda -i \partial_\sigma
\slashed{x}\, \lambda -i \mu \slashed{x}\, \Gamma_{1234}\,\lambda] 
\label{QLK}
\ee
in the free ($g_s=0$) theory and where $\lambda$ is the conjugate
momentum to $\theta:=\theta^1 +i\theta^2$.
Similarly the bosonic generators $P^I$ and $J^{I+}$ are given by 
(at $\tau=0$)
\be
P^I=\int d\sigma p^I\qquad J^{+I}=\ft{1}{2\pi\alpha'} \int d\sigma
x^I\, .
\label{PLK}
\ee
The relevant (anti)-commutators of the
plane-wave supersymmetry algebra are
\cite{Blau:2001ne,Metsaev:2001bj}
\begin{subequations}\label{susyalgebra}
\begin{align}
[H,P^I] &= i \, \mu^2\, J^{+I} \label{KomHP}\\
[H,Q^+] &= \mu\, \Gamma_{1234}\, Q^+ \\
[P^I,Q^-] &= \mu\, \Gamma_{1234}\, \Gamma^I\, Q^-\\
\{Q^-_a,\bar Q^-_b\}&=2\delta_{ab}\, H - i\mu\,
(\Gamma_{ij}\, \Gamma_{1234})_{ab}\, J^{ij} + i\mu\,
(\Gamma_{i'j'}\, \Gamma_{1234})_{ab}\, J^{i'j'}\, . \label{KomQQ}
\end{align} 
\end{subequations}
plus the remaining, unmodified super-Poincare algebra (anti)-commutators, 
with all commutators involving $J^{+-}$ and $J^{I-}$ omitted.
The free ($g_s=0$) string generators given in \eqn{HLK2}, \eqn{QLK}
and \eqn{PLK} obey these (anti)-commutation relations.

Conventionally, string interactions are introduced in light-cone
gauge quantization via vertex operators \cite{GSW}. These are constructed
by demanding covariant transformation properties under 
supersymmetry transformations. In flat space they then take a universal
structure of the form
\be
\hat V(k^+,k^I) = [\,\mbox{polarization dependent terms}\,]\,\cdot\, 
e^{i\, [k^+\, \hat x^-(\sigma) + k^I\, \hat x^I(\sigma)]}
\label{vertop}
\ee
corresponding to a string excitation with null momentum $k^+$ and
transverse momentum $k^I$. To compute an $N$-particle scattering-amplitude
one evaluates the light-cone string theory path-integral 
with $N$ vertex operator insertions. In the light-cone gauge 
$\hat x^-(\sigma)$
is given in terms of a quadratic function of the transverse degrees
of freedom $\hat x^I(\sigma)$, the analogue of \eqn{vircstr}, introducing
quadratic terms into the exponential of \eqn{vertop}. This leads to
technical problems in the path integral evaluation of the correlators.
A standard trick \cite{GSW} to circumvent this problem 
for not too many external particles lies in going  to a Lorentz
frame where $k^+=0$. Then the amplitude may be calculated 
easily and the obtained result is simply 
covariantized by making the replacement
$k^I\to k^\mu$ \cite{GSW}. In the plane-wave background 
two complications for such a procedure arise: First of all 
the transverse momentum $k^I$ is
not a valid quantum number any longer which could label asymptotic 
states: $P^I$ does not commute with the Hamiltonian \eqn{KomHP}.
Moreover the absence of the $J^{i-}$ generators forbids one to
move to a Lorentz-frame with vanishing $P^+$ momentum.
Hence this approach seems to fail.

\begin{figure}[t]
\begin{center}
\begin{tabular}{c}
\epsfxsize=6cm\epsfbox{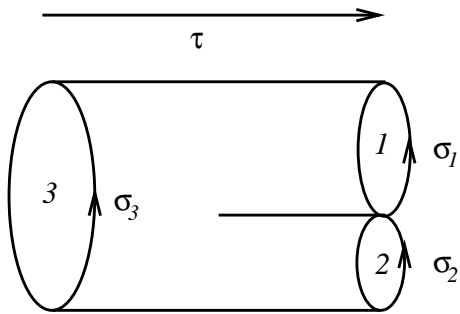}
\end{tabular}
\end{center}
\caption{The worldsheet of the three-string interaction vertex $\widehat H_3$.}
\label{3stringfig}
\end{figure}

As an alternative approach light-cone string field theory suggests
itself \cite{Mandelstam}\footnote{This subject is also developed
in chapter 11 of \cite{GSW}.}. Here one works in a multi-string Hilbert space 
built upon the multi-string vacuum:
\be
||0\rangle
 = |0\rangle_{(1)} \otimes|0\rangle_{(2)} \otimes|0\rangle_{(3)} 
\otimes\ldots
\label{StringFock}
\ee
upon which one acts with the creation operators $a^\dagger_{n(r)}$ and
$b^\dagger_{n(r)}$ of the individual strings,
where $r$ labels the corresponding
string-number. The string
field theory Hamiltonian acting in the multi-string Hilbert space
is then defined as an infinite  series in the string
coupling constant $g_s$
\be
\widehat H := \sum_{r} \widehat H_{2(r)} + g_s\, \widehat H_3 +
g_s^2\, \widehat H_4 + \ldots
\ee
The leading term $\widehat H_{2(r)}$ is the free string Hamiltonian
\eqn{HLK2} of the $r$'th string. Summing over $r$ yields the
free piece of the string Hamiltonian which conserves the number of
strings. The first interaction term $\widehat H_3$
corresponds to the three-string interaction vertex describing
a string splitting and joining process as depicted in figure 
\ref{3stringfig}. A general $n$-string state is
mapped by $\widehat H_3$ to the sum of an $(n+1)$-string 
and an $(n-1)$-string state.

What determines the structure of the interacting pieces of the
Hamiltonian $\widehat H_{n>2}$? In the simplest case
of the bosonic string in a flat background the structure of
$\widehat H_3$ is determined by demanding worldsheet continuity:
In the interaction process the two worldsheets 
of string 1 and 2 are smoothly glued together 
to form string 3. This is formally
achieved by the delta-functional expression
\be
\Delta[(X_{(1)}(\sigma_1)+X_{(2)}(\sigma_2)-X_{(3)}(\sigma_3)]\, .
\label{wscont1}
\ee
In practice such a delta-functional is represented by an infinite
product of individual delta-functions for each Fourier-mode of the
three $X_{(r)}(\sigma_r)$ involved. 
In light-cone string theory the total string length 
given by $\alpha'\, p^+$ is conserved in an interaction, which
is pictorially obvious from figure \ref{3stringfig} and an algebraic
consequence of the commutator relation $[H_3, P^+]=0$. This requirements
yields the additional delta-function contribution
$
\delta(p^+_{(1)}+p^+_{(2)}-p^+_{(3)})
$
to $\widehat H_3$. So 
\be
\widehat H_3^{\textindex{bosonic}}= \delta(p^+_{(1)}+p^+_{(2)}-p^+_{(3)})\,
\Delta[(X_{(1)}(\sigma_1)+X_{(2)}(\sigma_2)-X_{(3)}(\sigma_3)]\; .
\label{H3bosdef}
\ee
It is convenient to represent the operator
$\widehat H_3$ as a state in a 3-string Hilbert space via
\be
\langle\, \phi(3)\,|\, \widehat H_3\, |\,\phi(1)\rangle\, |\,\phi(2)\,\rangle
=: \langle\, \phi(3)\,|\, \langle \,\phi(2)\,|\, \langle\, \phi(1)\,|\, H_3
\,\rangle \, .
\label{h3state}
\ee
It can be shown that the bosonic cubic interaction vertex is
represented by the coherent three-string state \cite{Mandelstam}
\be
|H_3^{\textindex{bosonic}}\rangle 
\sim \exp\{ \ft 1 2 \sum_{r,s=1}^3 \sum_{m,n\in\Z}
a^\dagger_{m(s)}\, \bar N_{mn}^{rs}\,    a^\dagger_{n(r)}\,\}\,
|0\rangle_{123}
\ee
where $\bar N_{mn}^{rs}$ are known as the  
Neumann-matrices which are $c$-number valued entities
following from \eqn{H3bosdef}. Moreover we have defined
$|0\rangle_{123}=|0\rangle_{(1)}\otimes|0\rangle_{(2)}\otimes
|0\rangle_{(3)}$.

We are interested in the generalization of this to the
light-cone superstring in a
plane-wave background. Light-cone superstring field theory in
a flat background was
developed in the eighties by Green, Schwarz and Brink
\cite{Brinkxxx}. This has been generalized in a number of papers
to the plane-wave background 
\cite{Spradlin:2002ar,Spradlin:2002rv,Pankiewicz:2002tg,Chu:2002eu,Schwarz:2002bc,Pankiewicz:2002gs,Chu:2002wj,He:2002zu,Pankiewicz:2003kj}, 
which we will review in the following. 
The construction principle in the supersymmetric case
is the requirement that the superalgebra of \eqn{susyalgebra}
should be realized in the full interacting string field theory,
i.e.~in the presence of $g_s$ corrections\footnote{The
principle of worldsheet continuity in the bosonic case may also
be viewed as a preservation of the Poincar\/e algebra under $g_s$
corrections.}. One hopes that this uniquely determines the
form of the higher $g_s$ corrections to all generators of the
plane-wave superalgebra. In particular these generators may be divided
into two sets reflecting the occurrence of $g_s$ corrections: 
\begin{itemize}
\item {\bf Kinematical generators:} $P^+$, $P^I$, $J^{+I}$, $J^{ij}$,
$J^{i'j'}$, $Q^+$ \\
These do not contain derivatives in $\partial_{x^+}=\partial_\tau$,
they act at fixed light-cone time. These operators are not corrected
by interactions and preserve the string-number.
\item {\bf Dynamical generators:} $H$, $Q^-$\\
These operators are corrected by $g_s$ terms and are the objects
we are after. They create and destroy strings.
\end{itemize}
The requirement that the supersymmetry algebra 
\eqn{susyalgebra}
is preserved under
interactions now leads to two sets of constraints: {\sl Kinematical}
constraints arise from the (anti)-commutation relations of dynamical
with kinematical operators, $[D,K]=K$, and {\sl dynamical}
constraints arise from the (anti)-commutation relations of dynamical
operators alone, $[D,D]=D+K$. 

Let us first study the kinematical constraints. In the bosonic
sector the commutation relation of the full Hamiltonian with the
kinematical generators $P^I$ reads $[H,P^I]=i \mu^2\, J^{+I}$.
As only $H$ is corrected by interactions one immediately concludes
that the interaction pieces $H_{n>2}$ {\sl commute} with $P^I$
and hence conserve transverse momentum as in flat space
\be
[H_3,P^I]=0\quad \Longrightarrow \quad 
\sum_{r=1}^3 \widehat P_{(r)}(\sigma_r)
\, |H_3\rangle =0 \, .
\ee
Similarly $[H,J^{+i}]=-i P^I$ leads to
\be
[H_3,J^{+I}]=0\quad \Longrightarrow \quad 
\Bigl (  \widehat X_{(3)}(\sigma_3) - \widehat X_{(1)}(\sigma_1)- 
\widehat X_{(2)}(\sigma_2)
\Bigr )
\, |H_3\rangle =0 \, ,
\ee
the worldsheet continuity condition. In the fermionic sector the
commutator $[H,Q^+]=\mu\, \Gamma_{1234}\, Q^+$ yields
the kinematical constraint
\be
[H_3,Q^+]=0 \quad \Longrightarrow \quad 
\sum_{r=1}^3 \lambda_{(r)}(\sigma_r)
\, |H_3\rangle =0 \, .
\ee
with $\lambda_{(r)}$ being the conjugate momentum to $\theta_{(r)}$.
These kinematical constraints can be solved through the following 
state \cite{Spradlin:2002ar,Pankiewicz:2002gs}
\be
|H_3\rangle \sim |E_a\rangle\, |E_b\rangle\, \delta(p^+_{(3)}-
p^+_{(1)}-p^+_{(2)})\, =: |V\rangle
\label{Vdef}
\ee
where
\begin{subequations}
\begin{align}
|E_a\rangle &=\exp\{ \ft 1 2 \sum_{r,s=1}^3 \sum_{m,n\in\Z}
a^\dagger_{m(r)}\, \bar N_{mn}^{rs}(\mu)\,    a^\dagger_{n(s)}\,\}\,
|0\rangle_{123}\\
|E_b\rangle &= \exp\{\sum_{r,s=1}^3 \sum_{m,n>0}
b^\dagger_{-m(r)}\, Q_{mn}^{rs}(\mu)\,    b^\dagger_{n(s)}
-\sqrt 2\, \Lambda\sum_{m>0}Q^r_m\, b^\dagger_{-m(r)}
\,\}\,
|E^0_b\rangle
\end{align}
\end{subequations}
Here $\bar N^{rs}_{mn}(\mu)$ are the generalized Neumann matrices
for the plane-wave background, which are explicitly known functions
of $\mu$ \cite{He:2002zu}. Similarly $Q^{rs}_{mn}$ and $Q^r_m$ may
be expressed in terms of $\bar N^{rs}_{mn}(\mu)$ \cite{Pankiewicz:2002gs}.
Moreover $\Lambda$ is a linear function in the fermion zero modes
$\Lambda=\alpha'(p^+_{(1)}\, \lambda_{0(2)}
-p^+_{(2)}\, \lambda_{0(1)})$ and 
$|E_b^0\rangle = \prod_{a=1}^8\Bigl [\sum_{r=1}^3\lambda_{0(r)}^a\Bigr ]
\, |0\rangle_{123}$ is the pure zero-mode part of the fermion 
vertex\footnote{Note that there is a
subtlety in the fermionic zero mode structure of $|0\rangle_{123}$
related to its $\Z_2$ parity under the exchange of the two transverse $\R^4$
spaces. 
The choice of \cite{Chu:2002eu,Chu:2002wj,Pankiewicz:2003kj}
differs from the one we employ in these lectures
used in \cite{Spradlin:2002ar,Spradlin:2002rv,Pankiewicz:2002tg,Schwarz:2002bc,Pankiewicz:2002gs,He:2002zu}. Yet, in \cite{Pankiewicz:2003kj} evidence is presented
that the two choices are physically equivalent.
\label{fussnote}}.

As we see the emerging structures become
rather complicated, however, this is not the end of the story as we
still need to implement the dynamical constraint to completely
fix $|H_3\rangle$. For this the central anti-commutator is
\eqn{KomQQ} which one rewrites by defining the linear combinations
of supercharges ($\eta=e^{i\pi/4}$)
\be
\sqrt 2\, \eta\,  Q := Q^-+ i \bar Q^- \qquad
\sqrt 2\, \bar\eta\,  \tilde Q := Q^-- i \bar Q^- 
\label{Qnew}
\ee
to obtain from \eqn{KomQQ}
\begin{align}
\{ Q_a,\tilde Q_b\}&=
-i\mu(\Gamma_{ij}\, \Gamma_{1234})_{ab}\, J^{ij}
+i\mu(\Gamma_{i'j'}\, \Gamma_{1234})_{ab}\, J^{i'j'}\nn\\
\{ Q_a, Q_b\} &= \{\tilde Q_a,\tilde Q_b\} = 2\delta_{ab}\, H \, .
\label{KomQQ2}
\end{align}
As the angular momentum generators are kinematical  
the following dynamical constraint equations follow from this
\begin{align}
\sum_{r=1}^3 \Bigl (\, \tilde Q_{2\, a(r)}\, |Q_{3\, b}\rangle +
 Q_{2\, b(r)}\, |\tilde Q_{3\, a}\rangle\, \Bigr ) & = 0\nn\\
\sum_{r=1}^3 \Bigl (\, Q_{2\, a(r)}\, |Q_{3\, b}\rangle +
 Q_{2\, b(r)}\, |Q_{3\, a}\rangle \, \Bigr ) & 
= 2\delta_{ab}\, |H_3\rangle\nn\\
\sum_{r=1}^3 \Bigl(\, \tilde Q_{2\, a(r)}\, |\tilde Q_{3\, b}\rangle +
\tilde Q_{2\, b(r)}\, |\tilde Q_{3\, a}\rangle \,\Bigr ) & 
= 2\delta_{ab}\, |H_3\rangle\; . \label{dynconstr}
\end{align}
Here we have introduced the states $|Q_{3\, a}\rangle$ and
$|\tilde Q_{3\, a}\rangle$ corresponding
to the ${\cal O}(g_s)$ corrections of the dynamical supercharges 
\eqn{Qnew} in 
analogy to \eqn{h3state}. These constraints will be satisfied by
augmenting the kinematical part of the vertex $|V\rangle$ of
\eqn{Vdef} with prefactors $h_3, q_3, \tilde q_3$
being polynomials in the oscillators
$a^{\dagger\, I}_{m(r)}$ and $b^{\dagger}_{m(r)}$ 
\be
|H_3\rangle = h_3\, |V\rangle \qquad
|Q_3\rangle = q_3\, |V\rangle \qquad
|\tilde Q_3\rangle = \tilde q_3\, |V\rangle\, .
\ee
These prefactors are constructed in such a fashion that they
commute with the kinematical constraints in order to not spoil
the achievements of $|V\rangle$ and at the same time 
enforce the dynamical constraints \eqn{dynconstr}. It turns out
that there are three functions $K^I$, $\tilde K^I$ and $Y_a$ which
have the property of commuting with the kinematical constraints.
They are linear functions in the oscillators
\begin{align}
K^I&=\sum_{r=1}^3(
\sum_{n\geq 0}  F_{n(r)}\, a^{\dagger\, I}_{n(r)}
+\sum_{n> 0} F_{-n(r)}\, a^{\dagger\, I}_{-n(r)}) \nn\\
\tilde K^I&=\sum_{r=1}^3( \sum_{n\geq 0} ( F_{n(r)}\, a^{\dagger\, I}_{n(r)}
-\sum_{n> 0}F_{-n(r)}\, a^{\dagger\, I}_{-n(r)}) \nn\\
Y&=\sum_{r=1}^2 G_{0\, (r)}\, \lambda_{0\, (r)} + \sum_{r=1}^3
\sum_{m>0}G_{m(r)}\, b^{\dagger}_{m(r)}\, .
\end{align}
The explicit form of the functions $F_n$ and $G_n$ entering in the above 
may be found in \cite{Pankiewicz:2003pg}.
They obey relations of the form
\be
-\sqrt{2}\eta\kappa\,
\sum_{r=1}^3Q_{2\, (r)} \, |V\rangle = K^I\, \Gamma^I\, Y\, |V\rangle
\qquad
-\sqrt2 \bar \eta\kappa
\sum_{r=1}^3\tilde Q_{2\, (r)} \, |V\rangle = \tilde
K^I\, \Gamma^I\, Y\, |V\rangle \, ,
\ee
with $\kappa:=\alpha'^2p^+_{(1)}p^+_{(2)}p^+_{(3)}$.
This in view of \eqn{dynconstr} 
makes them ideal building blocks for an ansatz for the prefactors
$h_3$, $q_3$ and $\tilde q_3$. Utilizing these tools 
the final form of the dynamical supercharges and three-string
vertex was derived in  \cite{Spradlin:2002rv,Pankiewicz:2002tg}
\begin{align}
&|H_3\rangle = \Bigl ( \tilde K^I\, K^J - \mu\,\kappa\, \delta^{IJ}\Bigr )\,
v_{IJ}(Y)\, |V\rangle,\nn\\
&|Q_{3\, a}\rangle = \tilde K^I\, s^I_a(Y)\, |V\rangle, 
\qquad|\tilde Q_{3\, a}\rangle =  K^I\, \tilde s^I_a(Y)\, |V\rangle
\; , \label{SFTresult}
\end{align}
where 
\bea
v_{IJ}&=& \delta_{IJ} - \ft {i} {2\kappa} (Y\Gamma_{IJ} Y) + 
\ft {1}{4!\,\kappa^2} (Y\Gamma_{IK}Y)(Y\Gamma_{JK}Y)\nn\\&& 
-\ft {1}{2\cdot 6!\,\kappa^3}
(\Gamma_{IJ})_{ab}\, \epsilon^{ab}{}_{cdefgh} Y^c\ldots Y^h
+\ft {1}{8!\,\kappa^4}\delta_{IJ}\, \epsilon_{abcdefgh} Y^a\ldots Y^h
\nn\\
s^I_a&=&-i2\sqrt 2 \Bigl ( \eta (Y\Gamma^I)_a 
-\ft {\bar\eta}{3!\, \kappa}(Y\Gamma_{IJ}Y)
 (Y\Gamma_J)_a\, -\ft{\eta}{6!\, \kappa^2}(\Gamma_{IJ})_{bc}(\Gamma_J)_{da}
\epsilon^{bcd}{}_{efghi} Y^e\ldots Y^i\nn\\
&&
+\ft{\bar\eta}{7!\,\kappa^3}(\Gamma_I)_{ba}\, 
\epsilon^b{}_{cdefghi} Y^c\ldots Y^i
\Bigr )
\eea
and a similarly intricate expression for $\tilde s^I_a$. As a matter
of fact the overall normalization of $|H_3\rangle$, $|Q_{3\, a}\rangle$
and $|\tilde Q_{3\, a}\rangle$ is not fixed by closing the
plane-wave superalgebra. It is a priori an arbitrary function
of $\mu$, $\alpha'$ and the $p^+_{(r)}$'s. 
This ambiguity is due to the missing dynamical
Poincar\'e generators $J^{I-}$ in the algebra which 
fixes these normalizations in flat space. 

\subsection{Comparison to gauge theory at non-zero $g_2$}

Having obtained these results one
is in the position to explicitly evaluate matrix elements of
the three string interaction vertex and compare the result
with the Super Yang-Mills dilatation operator. Let us 
look at matrix elements involving  the first stringy bosonic excitation
$|n;p^+\rangle := a^{\dagger\, 1}_n\, a^{\dagger\, 2}_{-n}\, |0;p^+\rangle$
to find
\be
g_s\, \langle n; p^+|\, \widehat H_3\, | m;r\,p^+\rangle\, | 0;(1-r)\, p^+\rangle
= \mu\, g_s\, \lambda'\, (1-r)\, \frac{\sin^2(n\, \pi \,r)}{2\pi^2}
+{\cal O}(1/\mu)\, ,
\label{res1}
\ee
with $r\in[0,1]$ parameterizing the fraction of $p^+$ momentum and
where we have displayed only the leading term in the $\mu\to \infty$ limit.

Note the complementarity of the string theory and gauge theory 
perturbation expansions:
On the string theory side one has an expansion in $g_s$ with
a non-trivial (in principle known) dependence on $\mu$ for
every term. Perturbative
gauge theory on the other hand is organized in an expansion in 
$\lambda'\sim 1/\mu$ with a non-trivial (in principle known) dependence
on $g_2$ for every term in the quantum-loop expansion.
 
If one now naively compares the string result \eqn{res1}
to the ``corresponding'' 
one-loop matrix
element of the Super Yang-Mills dilatation operator of \eqn{nHmr} 
\be
\lambda'\, g_s\, \langle n| \hat H_{\rm int}|m,r\rangle = 
\lambda'\, g_s\,
\frac{r}{(1-r)^{1/2}}
\, \frac{4n}{n-\ft{m}{1-r}}\, \sin^2 (\pi n r)\, 
\ee
one
sees that they do not agree! However, this should not be too surprising, 
as we are dealing with an operator correspondence here
\be
\frac{\widehat H_{\rm  string}}{\mu} \entspricht {\cal D}- J\cdot \unit\, ,
\label{oprel}
\ee
relating two operators which act in two distinct Hilbert-spaces.
Hence if one wants to compare matrix elements of $\widehat H_{\rm string}$
and ${\cal D}$ one also has to provide  an isomorphism between
the string and the gauge theory Hilbert-spaces
\cite{Gross:2002mh,Gomis:2002wi}.

An immediate consequence of the relation \eqn{oprel} irrespective of the
isomorphism question is the agreement of the 
operator eigenvalues. On the gauge theory side we have seen how to
compute them up to genus two in eq.~\eqn{gtg2}. 
In string field theory the corresponding
calculation has been performed in \cite{Roiban:2002xr,Gomis:2003kj}
up to genus one.
To compute the mass shift of the single-string state 
$|n\rangle := a^{\dagger\, 1}_n\, a^{\dagger\, 2}_{-n}\, |0\rangle$
one uses non-degenerate perturbation theory, in analogy to the
gauge theory computation. At leading order (${\cal O}(g_s^2)$) 
the correction to the
eigenvalue of $|n\rangle$ comes from a one-loop diagram involving
$\widehat H_3$ and a contact term involving $\widehat H_4$, of which
only the $\{Q_{3\, a},Q_{3\, a}\}$ piece contributes due to the
necessity of having intermediate two-string states only\footnote{
The term $\{Q_{2\, a},Q_{4\, a}\}$ does not contribute as $Q_4$
takes a single-string state to a triple-string state, while $Q_2$
preserves the string number.}

\be
\delta E_{|n\rangle}^{(1)}=g_2{}^2\, \sum_{
|\alpha\rangle\, \in \, {\rm 2-string\, states}} 
\Bigl \{
\frac{\langle n| \widehat H_3 |\alpha\rangle\,
\langle \alpha| \widehat H_3| n\rangle}{E_{n}^{(0)}-E_\alpha^{(0)}} +
\ft 1 8 
\langle n | Q_{3\, a} | \alpha\rangle\, \langle 
\alpha| Q_{3\, a}|n\rangle\, \Bigr \}\, .
\label{hier}
\ee
From the final results sketched in \eqn{SFTresult} it is now possible to
evaluate the above mass shift. At this point, however, it has only been
possible to technically handle this computation by restricting 
the intermediate two-string channel $|\alpha\rangle$ to 
the ``impurity-conserving'' sector, i.e.~the sector 
containing precisely two oscillators 
acting on the two-string vacuum. This 
truncation of the computation
parallels the computation performed on the gauge theory side in
\eqn{SYMg1}. It is far from obvious at this point why such a 
truncation should be consistent, as there are contributions to the
``impurity-non-conserving'' channel, e.g.~the matrix element
\be
\langle 0;r\, p^+|\, \langle 0;(1-r)\, p^+| \widehat H_3\,
|n\rangle \neq 0
\ee
does not vanish and moreover is proportional to $\sqrt{\lambda'}$
\cite{Klebanov:2002mp}.
It remains to be seen whether it is consistent to suppress
the ``impurity-non-conserving'' channel, presently the contributions
from this sector appear to diverge \cite{Roiban:2002xr}, which may be due to
an order of limits problem.

Once one is willing to perform this restriction, however,  the
genus one correction to the mass shift
\be
\delta E_{|n\rangle}^{(1)}=\frac{\mu\, g_2{}^2\, \lambda'}{4\pi^2}
\, ( \frac{1}{12}+\frac{35}{32\, n^2\, \pi^2}) 
\ee
follows, in perfect agreement with the gauge theory result
\eqn{oneline} - a further strong confirmation of the plane-wave
string/Super Yang-Mills correspondence.

Let us now discuss the issue raised above concerning the existence of an
isomorphism relating the string field theory and the gauge
theory Hilbert-spaces. The natural bases we have been working
with are the multi-string Fock-space of \eqn{StringFock}
and the multi-trace states of \eqn{multi-trace-ops}. Both
come with a natural scalar product: The string field theory
one is obvious and induced by the single-string scalar-product, 
with orthogonality on single and multi-string states
\be
\langle s_\alpha|s_\beta\rangle = \delta_{\alpha\beta}\, .
\ee
On the gauge theory side it appears natural 
to use the free, tree-level ($\lambda'=0$) two-point
functions as the scalar product of multi-trace operators,
\be
\vev{\cO_\alpha(x)\, \bar\cO_\beta(0)}_{\rm tree-level}= 
\left (\frac{\gym^2}{8\pi^2|x|^2} \right )^{J+2}\,
S_{\alpha\beta} \, ,
\qquad
\Rightarrow \qquad
\langle \cO_{\alpha}| \cO_{\beta}\rangle := S_{\alpha\beta}\, .
\label{GTSP}
\ee
But as we have discussed at length in section four,
$S_{\alpha\beta}$ is not diagonal at higher orders in $g_2$
on single and multi-trace
states, due to the non-planar hopping graphs of figure \ref{Fig3}
on page \pageref{Fig3} 
\be
S_{\alpha\beta}=\delta_{\alpha\beta} + g_2\, S^{(1)}_{\alpha\beta}
+ g_2{}^2\, S^{(2)}_{\alpha\beta}
+\ldots
\label{Sexp}
\ee
When $g_2=0$ the string field theory and gauge theory bases coincide:
single string-states are to be identified with single-trace operators
according to the BMN dictionary of section 4.1. Similarly multi-string
states are identified with multi-trace operators. Taking into 
account corrections in $g_2$ the Super Yang-Mills trace-operators
start mixing, whereas the string-states remain orthogonal. In
order to remedy this situation one
seeks a transformation matrix $U_{\alpha\beta}$
of the gauge theory states which diagonalizes $S_{\alpha\beta}$, 
\footnote{$\Gamma_{\alpha\beta}$ {\sl only} diagonalizes
the tree-level two-point function, not to be confused
with $V_{\alpha A}$ of \eqn{VaA} diagonalizing the all loop
two point-function.}
\be
|\tilde \cO_\alpha\rangle = \Gamma_{\alpha\beta}\, |\cO_\beta\rangle \qquad
\Rightarrow \qquad
\Gamma\, S\, \Gamma^{\dagger} = \unit\, .
\label{Grossprop}
\ee
Clearly $\Gamma$ is given in terms of an expansion in $g_2$
\be
\Gamma= \unit + g_2\, \Gamma^{(1)}+ g_2{}^2\, \Gamma^{(2)} + \ldots \, .
\ee
In the new  $|\tilde\cO_\alpha\rangle$ basis the Super 
Yang-Mills dilatation operator $\tilde {\cal D}$ is
directly given by the one-loop piece 
$\tilde T_{\alpha\beta}= (U\, T\, U^\dagger)_{\alpha\beta}$ of
the two point function \eqn{Gen2PtFct} and its matrix elements
should directly correspond to the matrix elements of the string field
theory Hamiltonian $\widehat H_{\rm string}$ 
\cite{Gross:2002mh,Gomis:2002wi}
\be
\tilde {\cal D}_{\alpha\beta} - J\cdot \delta_{\alpha\beta} 
= \underbrace{\tilde T_{\alpha\beta}}_{(\Gamma\, T\, 
\Gamma^\dagger)_{\alpha\beta}} - J\cdot \delta_{\alpha\beta} \quad
\entspricht\quad  \langle s_\alpha| \widehat H_{\rm string}| 
s_\beta\rangle \, .
\label{comp}
\ee
However, there is a crucial problem with this proposal: The 
transformation matrix $\Gamma_{\alpha\beta}$ is not unique:
As a matter of fact the Hermitian matrix $S$ is diagonalized by
\be
\Gamma := U\cdot S^{-1/2}
\label{gammadef}
\ee
with $U$ an arbitrary unitary matrix.
But the form of $U$ will affect the
matrix elements of the dilatation operator in \eqn{comp}.
In \cite{Gross:2002mh} it was proposed to fix this
ambiguity by simply {\sl demanding} agreement of three-string 
vertex and gauge theory dilatation operator matrix elements. 
This situation is not satisfactory, as one
looses the predictive power of the duality conjecture.
The non-trivial statement would then solely lie in the mere existence
of a transformation matrix $\Gamma$, which diagonalizes
$S$ and lets \eqn{comp} hold. This statement is equivalent to the match
of eigenvalues of the operators $\widehat H_{\rm string}$ and
${\cal D}$. In \cite{Gomis:2002wi} it was noted that
the agreement of the string and gauge theory
matrix elements  precisely occurs if $\Gamma$ is a symmetric, real
matrix, i.e.~if $U=\unit$. The authors of \cite{Gomis:2002wi} 
proposed to introduce this
as a fundamental property of $\Gamma$ and to diagonalize with
$S^{-1/2}$.
Using this prescription to date all comparisons of matrix elements of
the string field theory Hamiltonian with the corresponding
matrix elements of the Super Yang-Mills dilatation operator
have been shown to agree at the one-loop
level\footnote{Note, however, the discrepancies of a factor of 2
mentioned at order $(\lambda')^2$ in \cite{Spradlin:2003bw}.}.  
It is clearly highly desirable to
obtain an understanding why this particular change of basis is
singled out. 

The structure of the tree-level mixing matrix $S_{\alpha\beta}$
is very intriguing. There are good indications that the
full $g_2$ expansion of \eqn{Sexp} simply exponentiates, i.e.
\be
S = \exp[ \, g_2\, S^{(1)}] \,.
\label{Sasexp}
\ee
This has been shown to be the case in the ground state sector
$\Tr Z^J$ to all orders in $g_s$ and in the two-impurity sector to order
$g_2{}^2$ in \cite{Vaman:2002ka}. Moreover, the conjugation relation 
of the effective quantum mechanical Hamiltonian in 
\eqn{nonherm} has been shown to hold for this particular choice of 
$S$ in the space of two-impurity states in
\cite{Spradlin:2003bw}. The ansatz \eqn{Sasexp}
is motivated by a discretized
string model due to H. Verlinde and collaborators
\cite{stringbits,Vaman:2002ka,pearson} on which we shall briefly
comment. We have seen in section 4.2 how a
discretized string picture naturally emerges from the gauge
theoretic considerations, with $J$ corresponding to the number
of string ``bits''. In \cite{stringbits,Zhou:2002mi,Vaman:2002ka,pearson} 
a direct
discretization of the space-like component of the 
plane-wave superstring world-sheet in light-cone gauge
was studied, resulting in a quantum mechanical Hamiltonian 
of  $J$ constituent bits. String interactions weighted by $g_2$
can be included in this model and the diagonalization
of the Hamiltonian to order $g_2{}^2$ has been performed
in a hybrid string bit/string field theory computation in
\cite{pearson} finding agreement with the gauge theory 
prediction \eqn{oneline}. The string bit formalism lies 
``in-between'' the continuum string field theory 
and BMN gauge theory. The conjectured exponential structure of the
tree-level mixing matrix $S$ in \eqn{Sasexp} can be understood 
combinatorially in the string bit model where
$S^{(1)}$ is identified as a string bit 
permutation operator \cite{Vaman:2002ka}. In \cite{Spradlin:2003bw}
a nice direct connection of the effective BMN quantum mechanics discussed
in section 5.1 and the string bit model was demonstrated.
In \cite{Bellucci:2003qi} it was shown 
that the string bit model is plagued by 
the familiar problem of fermion doubling, in \cite{Danielsson:2003yc} a way
to evade this problem was pointed out.

Finally let us mention that there exists an alternative approach to
study string interactions from perturbative gauge theory in the
BMN limit based
on the collective field method \cite{jevicki}.

\subsection{Summary of the performed tests of the duality}

Let us give an account of the tests of the plane-wave
string/gauge theory duality performed so far organized by
the order of $g_2$ (genus counting parameter)
and $\lambda'$ (effective gauge theory quantum loop 
counting parameter) that was probed in the
operator relation \eqn{oprel}: 
\begin{itemize}
\item \mb$g_2{}^0\, \lambda'^n$\mn\\
On the string theory side the result is known to all orders in $n$, 
as discussed in \eqn{spectrum}. 
For two scalar impurities the gauge theory result
is known (based on certain assumptions) to all orders in $n$ as
well and agrees\cite{Santambrogio:2002sb}. Checks for $n=1$ exist for
scalar-vector impurities \cite{Gursoy:2002yy} and for two
vector impurities \cite{Klose:2003tw}. In \cite{Beisert:2002tn}
it is proved that {\sl all} two impurity (scalar, fermion, vector)
gauge theory excitation
are degenerate in scaling dimensions
with the two scalar impurity state to all orders
in $\lambda'$ and $g_2$.
\item \mb$g_2{}^1\, \lambda'^n$\mn\\
Here again the string theory result is known in principle to all
orders in $n$ from the explicit form of $\widehat H_3$. 
Tests for $n=1$ employing the map $\Gamma=S^{-1/2}$ 
of the gauge theory basis to the string theory basis have been performed
in the two impurity sector
for scalar impurities in \cite{Gomis:2002wi} and vector-scalar and
pure vector impurities in \cite{Georgiou:2003kt}. Tests for an
arbitrary number of scalar impurities were perfomed in
\cite{Georgiou:2003kt} at $n=1$.
\item \mb$g_2{}^{2}\, \lambda'$\mn\\
At order $g_2{}^2$ the eigenvalues of the two 
operators in \eqn{oprel} are compared,
which is a basis independent statement. The string theory result
for all two scalar impurities was computed in 
\cite{Roiban:2002xr,Gomis:2003kj} by truncating to the
impurity conserving channel as discussed in the previous section.
The result matches the gauge theory result of
\cite{Beisert:2002bb,Constable:2002vq}. Again due to \cite{Beisert:2002tn}
all two impurity excitations (bosonic and fermionic) are known to
have identical scaling dimensions.
\item \mb$g_2{}^{2}\, \lambda'^2$\mn\\
Here the two loop gauge theory analysis was carried out in
\cite{Beisert:2003tq} for two impurity states. 
The string theory result has been 
given in \cite{Spradlin:2003bw} and a {\sl disagreement}
by a factor of 2 was reported in a comparison of $\widehat H_3$ matrix elements
employing the map $\Gamma=S^{-1/2}$.
\item \mb$g_2{}^4\, \lambda'$\mn\\
Here only the gauge theory result for the eigenvalue of
the dilatation operator is known for two impurity states
\cite{Beisert:2002ff}
and non-degenerate perturbation theory breaks down \cite{Freedman:2003bh}.
\end{itemize}
Clearly the worrisome point is the mismatch at order $g_2{}^2\, \lambda'^2$
and should be clarified. Let us also note once more that in principle
in the two-impurity sector
the all orders result $g_2{}^m\, \lambda'^n$ in gauge theory
follows from the diagonalization of the effective quantum 
mechanical Hamiltonian $H_{\rm full}$ of \eqn{Hfull} -- if one is willing
to extrapolate the $\lambda'^2$ result to the square root expression.

\subsection{An alternative ansatz for plane-wave string field
theory}

We should mention that there exists an alternative proposal for the
construction of the three-string vertex due to Di Vecchia,
Petersen, Petrini, Russo and Tanzini
\cite{DiVecchia:2003yp} differing from the approach
discussed above. This approach departs from the one of
Spradlin, Volovich, Pankiewicz, Stefanski and others in 
the construction of the prefactors needed to satisfy the dynamical
constraint. The alternative ansatz of \cite{DiVecchia:2003yp} to solve
the dynamical constraints \eqn{dynconstr} is remarkably simple, 
\be
|H_3\rangle = \sum_{r=1}^3 \widehat H_{2\, (r)} |V\rangle \, , \qquad
|Q_{3\, a}\rangle = \sum_{r=1}^3 Q_{2\, a(r)}|V\rangle \, , \qquad
|\tilde Q_{3\, a}\rangle = \sum_{r=1}^3 \tilde Q_{2\, a(r)}
|V\rangle \, .
\label{aansatz}
\ee
the prefactors are produced by acting with the free Hamiltonian
and supercharges on the kinematical vertex 
$|V\rangle$\footnote{Here $|V\rangle$ is the kinematical vertex
built upon the vacuum state choice of 
\cite{Chu:2002eu,Chu:2002wj,Pankiewicz:2003kj} compare footnote
\ref{fussnote}.}. 
With this ansatz the dynamical constraints are automatically
fulfilled \eqn{dynconstr} as a consequence of the plane-wave superalgebra
\eqn{susyalgebra}\footnote{Note that the kinematical vertex $|V\rangle$
is annihilated by the angular-momenta $J^{ij}$ and $J^{i'j'}$, as it 
is $SO(4)\times SO(4)$ invariant.}. One then needs to check that this
alternative ansatz \eqn{aansatz} does not ruin the kinematical
constraints, which the authors of \cite{DiVecchia:2003yp} demonstrate.
In contradistinction to the vertex discussed in the previous
sections this
alternative vertex does {\sl not} lead to the flat space 
vertex in the $\mu\to 0$ limit. As a matter of fact the ansatz \eqn{aansatz} 
could equally well be written down for the flat background
string. It always has the property of yielding trivial 
on-shell amplitudes: The amplitude for energy conserving processes  
vanishes.

The new vertex does not satisfy the tests reported in the
previous sections upon making use of the transformation
matrix $\Gamma=S^{-1/2}$. The authors argue, however, that there
exists a different matrix $\Gamma$ in \eqn{gammadef} for which the gauge 
theory results can be made to agree with their
alternative proposal for $| H_3\rangle$.
In any case this is of no direct concern to 
the authors of \cite{DiVecchia:2003yp} as they propose to
abandon the operator correspondence \eqn{oprel} at the non-planar level
that we have been working with. Instead they propose to relate
their three-string vertex to the gauge theory three-point functions
in the {\sl naive} basis (for which the Super Yang-Mills
two-point functions will
{\sl not} be diagonal and the proposed dual operators do not have
a well defined scaling dimension). 
They demonstrate the validity of their proposal at leading order in 
$\lambda'$ through a number of checks.
Their proposed correspondence only makes
sense at tree-level (due to non-canonical coordinate dependencies
of the Super Yang-Mills three-point function at higher loops 
from not resolving the operator mixing) and it is
unclear how it could be generalized to higher point functions (due to
the non-existence of four point functions in the BMN limit 
\cite{Beisert:2002bb}). From our point of view this makes this
proposal deficient to the one discussed in these lectures.

But irrespective of the dual gauge theory matter 
the alternative ansatz indicates the non-uniqueness of the construction
of the three-string vertex from the plane-wave superalgebra.
This situation is very worrying and we hope that this problem
will be settled in future work.

\section{Outlook}

In these lectures we have developed the duality of type IIB string
theory in a maximally supersymmetric plane-wave background and the
BMN limit of ${\cal N}=4$, $d=4$ Super Yang-Mills at the free
and interacting string level. The key relation in this correspondence is
the identification of the string field Hamiltonian with the
dilatation operator of the Super Yang-Mills model minus the
$U(1)$ R-charge generator. The remarkable feature of this novel
string/gauge theory duality is its apparent perturbative structure
on both sides of the correspondence due to the emergence of
the effective coupling constants $\lambda'$ and $g_2$
in the BMN limit. This enabled us to perform
string computations by means of perturbative Yang-Mills
computations, which could even be further simplified to an effective
quantum mechanical description. 
A wealth of perturbative tests has been performed, probing higher
genus string theory and high quantum loop orders in Super Yang-Mills, 
as was summarized in these lectures. 
There are indications, however, of a breakdown
of this perturbative correspondence, which occur in impurity non-conserving
matrix elements of the string Hamiltonian of order $\sqrt{\lambda'}$.
These have no counterpart in perturbative Super Yang-Mills and
might be related to strong coupling effects in the gauge
theory which remain invisible in the effective weak coupling
computations we have performed\footnote{Note that in AdS/CFT strong
coupling predictions from string theory/supergravity typically
scale as $\sqrt{\lambda}$.}. More work along these lines is needed,
which could also lead to a deeper understanding of why the truncation of
the one-loop string field theory calculation to the ``impurity-conserving''
channel agrees with the perturbative gauge theoretic answer.

A further important question is what the effective quantum
mechanics describing the BMN sector of the ${\cal N}=4$
gauge theory really is. It was argued in \cite{BN} that it should be
given by a dimensional reduction of  ${\cal N}=4$ Super Yang-Mills
on a three sphere. This was studied in \cite{Kim:2003rz} where
it was shown that a consistent reduction on $S^3$ exists (and is
actually nothing but the plane-wave matrix model of 
\cite{Berenstein:2002jq} related to M-theory on a plane-wave).
However, this quantum mechanical model fails to reproduce
the eigenvalues of the BMN dilatation operator at two loop
order. So the nature of ``BMN quantum mechanics'' remains an enigma.
By working our way up perturbatively
we have seen first traces of an effective quantum mechanics for
BMN gauge theory in the two-impurity sector in our discussion in section 5.1.
But a deeper insight into its inner workings is still necessary.

This problem is closely related to the unsettled question of how
the holographic principle is realized in the plane-wave string/gauge
theory duality. For work along these lines see
\cite{pwholography}.

Instanton effects in the BMN gauge theory have not been addressed
so far. These should correspond to D-instantons of the dual
plane-wave superstring and it would be very interesting to see
whether their effects survive the BMN limit. 

Let us mention in closing that there are a number of topics
closely related to the material covered in these lectures,
which we could not address:
Corrections in $1/R^2$ to the plane-wave geometry from the
Penrose limit (compare \eqn{last}) have been considered.
In the free ($g_2=0$) string theory they give rise to
perturbative corrections in $1/R$ of the spectrum \cite{semicl2} and
these corrections have been reported to agree with planar $1/J$
corrections of the scaling dimensions on the gauge theory side \cite{1oR}.
A similar duality to the one studied here exists for open strings
in plane-wave backgrounds \cite{Berenstein:2002zw},
for which the interacting ($g_2\neq 0$) 
theory has been explored 
on the dual ${\cal N}=2$ $Sp(N)$ gauge theory side 
in \cite{Gomis:2003kb} and in open string field theory in
\cite{Chandrasekhar:2003fq}. Situations with less supersymmetries
have been studied as well, e.g.~by orbifolding the plane-wave background
resulting in ${\cal N}=2$ $[U(N)]^M$ quiver gauge theory
\cite{Kim:2002fp}. Moreover there is a lot of work on D-branes of the 
maximally supersymmetric plane-wave superstring \cite{Dbranes} 
which we did not discuss and first steps in
exploring their gauge theory duals have been undertaken in
\cite{Dbraneduals}.

\bigskip\bigskip\noindent
{\bf Acknowledgments}: \\ I would like to thank 
the organizers for an inspiring winter school in Turin, for inviting
me to present these lectures and for very
memorable dinners. Moreover, I wish to thank Niklas Beisert,
Charlotte Kristjansen, Gordon Semenoff and Matthias Staudacher
for a most pleasant collaboration on the gauge theory aspects
of this duality. Finally, I thank 
Gleb Arutyunov, Stefano Kovacs and Rodolfo Russo for useful discussions and
in particular Niklas Beisert, Ari Pankiewicz, Bogdan Stefanski
and Thomas Klose for very helpful comments on the manuscript.


\end{document}